\documentclass[namedreferences]{SolarPhysics}
\usepackage[optionalrh]{spr-sola-addons} 
\usepackage{graphicx}                    
\usepackage{color}                       
\usepackage{ulem}                      
\usepackage{url}                         
\usepackage{rotating}

\begin{document}

\begin{article}

\begin{opening}

\title{The Interaction of Successive Coronal Mass Ejections: A Review}
\author{No{\'e}~\surname{Lugaz}$^{1}$\sep Manuela~\surname{Temmer}$^{2}$\sep Yuming~\surname{Wang}$^{3}$\sep Charles~J.~\surname{Farrugia}$^{1}$}

\runningauthor{Lugaz et al.}
\runningtitle{CME-CME Interaction: a Review}

%
\institute{$^{1}$ Space Science Center and Department of Physics, University of New Hampshire, Durham, New Hampshire, USA.
                     email:~\url{noe.lugaz@unh.edu}; \url{charlie.farrugia@unh.edu}\newline                
                     $^{2}$ Institute of Physics, University of Graz, Universit\"atsplatz 5, A-8010, Graz, Austria. email:~\url{manuela.temmer@uni-graz.at}\newline
                     $^{3}$ School of Earth and Space Sciences, University of Science and Technology of China, Hefei 230026, China. email:~\url{ymwang@ustc.edu.cn} }
                     
\begin{abstract}
We present a review of the different aspects associated with the interaction of successive coronal mass ejections (CMEs) in the corona and inner heliosphere, focusing on the initiation of series of CMEs, their interaction in the heliosphere, the particle acceleration associated with successive CMEs, and the effect of compound events on Earth's magnetosphere. 
The two main mechanisms resulting in the eruption of series of CMEs are sympathetic eruptions, when one eruption triggers another, and homologous eruptions, when a series of similar eruptions originates from one active region. CME-CME interaction may also be associated with two unrelated eruptions. 
The interaction of successive CMEs has been observed remotely in coronagraphs (with the {\it Large Angle and Spectrometric Coronagraph Experiment} --LASCO--  since the early 2000s) and heliospheric imagers (since the late 2000s), and inferred from {\it in situ} measurements, starting with early measurements in the 1970s. The interaction of two or more CMEs is associated with complex phenomena, including magnetic reconnection, momentum exchange, the propagation of a fast magnetosonic shock through a magnetic ejecta, and changes in the CME expansion. The presence of a preceding CME a few hours before a fast eruption has been found to be connected with higher fluxes of solar energetic particles (SEPs), while CME-CME interaction occurring in the corona is often associated with unusual radio bursts, indicating electron acceleration. 
Higher suprathermal population, enhanced turbulence and wave activity, stronger shocks, and shock-shock or shock-CME interaction have been proposed as potential physical mechanisms to explain the observed associated SEP events. When measured {\it in situ}, CME-CME interaction may be associated with relatively well organized multiple-magnetic cloud events, instances of shocks propagating through a previous magnetic ejecta or more complex ejecta, when the characteristics of the individual eruptions cannot be easily distinguished. CME-CME interaction is associated with some of the most intense recorded geomagnetic storms. The compression of a CME by another and the propagation of a shock inside a magnetic ejecta can lead to extreme values of the southward magnetic field component, sometimes associated with large values of the dynamic pressure. This can result in intense geomagnetic storms, but also trigger substorms and large earthward motions of the magnetopause, potentially associated with changes in the outer radiation belts. 
Future {\it in situ} measurements in the inner heliosphere by {\it Solar Probe+} and {\it Solar Orbiter} may shed light on the evolution of CMEs as they interact, by providing opportunities for conjunction and evolutionary studies. 

\end{abstract}
\end{opening}

\section{Introduction}
Understanding coronal mass ejections (CMEs) is central to better grasp the complexities of the heliosphere, as they represent together with flares, the most intense phenomena in the Sun-Earth system. At the Sun, the exact cause(s) and trigger(s) of CME initiation are still a matter of debate (see review by \opencite{Chen:2011}), but it is well established that CMEs are one of the main ways for currents and magnetic energy to be released. CMEs typically consist of mostly closed magnetic field lines and carry mass and magnetic flux into the interplanetary (IP) space. Therefore, during times of high solar activity, CMEs highly structure the solar wind plasma and interplanetary magnetic field (IMF) characteristics in the IP space. 

CMEs play an important role in the heliospheric magnetic flux balance, by dragging magnetic field lines through the Alfv{\'e}n surface \cite{Owens:2006b,Schwadron:2010b}. CME-driven shocks are overwhelmingly thought to be the main accelerator of gradual solar energetic particles (SEPs) \cite{Kahler:1984,Reames:2013}. CMEs are also the primary drivers of intense geomagnetic storms at Earth \cite{Gonzalez:1987,Gosling:1991,Webb:2000,Zhang:2007}, and they are also associated with many of the strongest substorms \cite{Kamide:1998,Tsurutani:2015}, changes in Earth radiation belts \cite{Miyoshi:2005} and geomagnetically-induced currents (GICs) \cite{Huttunen:2008}. A recent review of CME research can be found in \inlinecite{Gopalswamy:2016}.

The rate of CMEs during the solar cycle is highly variable, ranging at the Sun from 2--3 CMEs {\it per} week in solar minimum to 5--6 CMEs {\it per} day in solar maximum. Some CME properties in the corona are now routinely measured by space-based coronagraphs such as the {\it Large Angle and Spectrometric Coronagraph Experiment} on board the {\it Solar and Heliospheric Observatory} (SOHO/LASCO: \opencite{Domingo:1995}, \opencite{Brueckner:1995}) and the {\it Solar-Terrestrial Relations Observatory} coronagraphs (STEREO/COR: \opencite{Kaiser:2008}). Catalogs such as the Coordinated Data Analysis Workshops (CDAW) CME catalog \cite{Yashiro:2008,Gopalswamy:2009b} report the CME speed, mass, acceleration, and angular width projected onto the plane-of-the-sky of the instruments. New catalogs such as such as the Heliospheric Cataloguing, Analysis and Techniques Service (HELCATS)\footnote{\url{http://www.helcats-fp7.eu}} based on STEREO/{\it Heliospheric Imager} (HI: \opencite{Eyles:2009}) observations give CME speed and direction in the IP space. CME properties near Earth are directly measured by spacecraft such as {\it Advanced Composition Explorer} (ACE), {\it Wind}, or {\it Deep Space Climate Observatory} (DSCOVR, operational since July 27, 2016). CME properties may be strongly influenced by their interaction with the solar wind and IMF. To first order, this interaction results in a deceleration of fast CMEs and an acceleration of slow CMEs \cite{Gopalswamy:2000,Vrsnak:2001,Cargill:2004,Liu:2013}, changes in the radial expansion rate of the magnetic ejecta \cite{Gulisano:2010,Poomvises:2010} and, sometimes, its deflection \cite{Wang:2014} and rotation \cite{Nieves:2012}. Adding to these broad tendencies, CME properties may  change even more drastically when they interact with corotating solar wind structures, such as fast wind streams and corotating interaction regions (CIRs) and with other CMEs. The interaction of a CME with a CIR has been studied both through numerical modeling as well data analysis \cite{Prise:2015,Winslow:2016}. 

Combining the CME frequency and their typical propagation time (3--4 days from Sun to Earth), they may be as few as two CMEs or as many as 20 in the 4$\pi$ sr between the Sun and the Earth, depending on the phase of the solar cycle. Assuming that a CME and its shock wave can be modeled as a cone of half-angle of 30$^\circ$, a CME occupies approximately $\pi$/4 sr. In solar maximum, interaction between unrelated successive CMEs is bound to happen; however, CME-CME interaction also happens regularly even in more quiet phases of the solar cycle. Solar observations often reveal that recurrent CMEs occur from the same active region, often associated with homologous flares \cite{Schmieder:1984,Svestka:1989}. On the other hand, sympathetic flares and CMEs may be an even more frequent cause of successive CMEs in relatively close angular and temporal separation ({\it i.e.} in optimal conditions for, at least, partial interaction). Early work based on coronagraphic observations \cite{Hansen:1974} and simulations \cite{Steinolfson:1982} discussed the possibility and consequences on the corona of successive, quasi-homologous eruptions. 

During their propagation from Sun to Earth, the interaction of successive CMEs may take a variety of forms: 1) the two CME-driven shock waves may interact without the ejecta interacting, 2) one shock wave may interact with a preceding magnetic ejecta, or 3) the successive magnetic ejecta may interact and/or reconnect. The fact that CMEs can interact on their way to Earth has been known for several decades now. Some of the early articles focused on the series of seven flares in 72 hours in early August 1972, and the associated three or four shock waves measured by {\it Pioneer 9}, {\it Prognoz} and the {\it the Interplanetary Monitoring Platform-5} (IPM-5) in the inner heliosphere and one shock wave measured by {\it Pioneer 10} at 2.2~AU \cite{Dryer:1976,Intriligator:1976,Ivanov:1982}.
For example, \inlinecite{Ivanov:1982} has a section focusing on ``shock waves from a series of flares'', where complex IP streams originating from compound shock waves and their interaction region are described. \inlinecite{Burlaga:1987} describe a variety of compound streams resulting from the interaction of a transient with another transient or with a solar wind stream. They discussed the interaction of two ejecta, one containing a magnetic cloud and one without, as well as three shock waves and noted that ``the compression of the magnetic cloud by shock S3 produced magnetic field strength up to 36~nT''.

The August 1972 series of eruptions resulted in a series of intense geomagnetic storms with the disturbed storm time (Dst) index peaking at $-154$~nT. \inlinecite{Tsurutani:1988} investigated the interplanetary origin of intense geomagnetic storms in the solar maximum of Solar Cycle 21, including cases related to the passage at Earth of a compound stream composed of multiple high-speed streams. \inlinecite{Burlaga:1987} studied the 3\,--\,4 April 1979 event associated with the interaction of two ejecta associated with an intense geomagnetic storm (Dst reached $-202$~nT) and discussed the relation between compound streams and large geomagnetic storms, finding that nine out of 17 large geomagnetic storms for which interplanetary data was available were associated with compound streams (this includes CIR-CME as well as CME-CME interaction). To explain this result, they noted that ``magnetic fields in ejecta can be amplified by the interaction with a shock and/or a fast flow and thereby cause a large geomagnetic storm'' and concluded that ``the interaction between two fast flows is in general a nonlinear process, and hence a compound stream is more than a linear superposition of its parts.''
Another multi-spacecraft study of compound streams measured in the late 1970s was performed by \inlinecite{Burlaga:1991}.

Once again, the 2\,--7\, August 1972 events revealed how series of flares and eruptions can result in extremely high level of SEPs \cite{Lin:1976}. \inlinecite{Sanderson:1992} discussed {\it Ulysses} measurements of a shock propagating inside a shock-driving magnetic cloud and the low level of energetic particles between the two shocks. This was explained as the magnetic cloud ``acting as a barrier delaying the onset of the high-energy protons from the second flare''. \inlinecite{Kallenrode:1993} discussed super-events associated with series of flares and CMEs.

\inlinecite{Vandas:1997} studied the interaction of a shock wave with a magnetic cloud using a 2.5-D magneto-hydrodynamical (MHD) simulation. This study illustrates the power of numerical simulations, as a case with an overtaking shock was compared with an identical case without an overtaking shock. The authors noted that the shock propagation results in a radial compression of the magnetic cloud, a change of its aspect ratio, acceleration as well as heating of the cloud. 

In the rest of the article, we focus primarily on developments about the causes and consequences of series of CMEs since 2000. The combination of LASCO imaging and {\it in situ} measurements at L1 from {\it Wind} and/or ACE since 1996 makes it possible to relate coronal observations with their {\it in situ} consequences and geomagnetic effects. The study of CME-CME interaction proliferated following the report of two CMEs interacting within the LASCO/C3 field-of-view and associated type II event \cite{Gopalswamy:2001} as well as the possible association of interacting CMEs with large SEP events \cite{Gopalswamy:2002}. Statistical surveys of geomagnetic storms and their interplanetary causes have become more routine during Solar Cycles 23 and 24 due to the reliability of L1 measurements; this has revealed how interacting CMEs may cause intense geomagnetic storms. In Solar Cycle 24, high spatial and temporal resolution observations by the {\it Solar Dynamics Observatory} (SDO: \opencite{Pesnell:2012}) have returned the study of sympathetic eruptions to central stage. The development of heliospheric imaging with the {\it Solar Mass Ejection Imager} (SMEI: \opencite{Eyles:2003}, \opencite{Jackson:2004}) and the HIs onboard STEREO have led to a large increase in the number of published cases of CME-CME interaction being remotely observed. Lastly, the development of large-scale time-dependent numerical simulations in the past 20 years have yielded new insights into the mechanisms resulting in the initiation of series of CMEs as well as the physical processes occurring during their propagation and interaction. 
This article is organized as follows. In Section~\ref{sec:initiation}, we discuss recent developments regarding the initiation of successive CMEs, including observations and numerical simulations of sympathetic and homologous CME initiation. In Section~\ref{sec:SEP}, we review observational and theoretical works focusing on the association of successive and interacting CMEs with large SEP events and with enhanced and unusual radio emissions. In Section~\ref{sec:helio}, we focus on the physical processes occurring during CME-CME interaction in the inner heliosphere, with insights gained from recent remote observations by SECCHI as well as by numerical simulations and the analysis of {\it in situ} measurements. In Section~\ref{sec:geo-effect}, we discuss how the complex ejecta resulting from CME-CME interaction may drive Earth's magnetosphere in unusual ways, often driving large geomagnetic storms, but also sometimes in weaker-than-expected storms. In Section~\ref{sec:conclusion}, we discuss what to expect in the upcoming decade with new observations closer to the Sun made possible by {\it Solar Probe+} and {\it Solar Orbiter} and conclude.

\section{Initiation of Successive CMEs}\label{sec:initiation}

\subsection{Trigger and Initiation of CMEs}

As the largest explosive phenomenon on the Sun, a typical CME carries about $10^{32}$ erg 
of energy~\cite{Vourlidas_etal_2000,Hudson_etal_2006} and $10^{21}$ Mx of magnetic 
flux~\cite{Dasso_etal_2005,Qiu_etal_2007,Wang_etal_2015} into IP space, associated
with reconfiguration of coronal magnetic fields in the CME source region. To support such 
large eruptions, the following is needed: (1) sufficient magnetic free energy, and (2) triggers and efficient energy conversion processes to release the free energy in a short timescale. The magnetic free energy
as well as helicity can be accumulated gradually via various ways, {\it e.g.}, flux 
emergence~(see, {\it e.g.} \opencite{Heyvaerts_etal_1977}, \opencite{Chen_Shibata_2000}), 
shearing/rotational motion~(see, {\it e.g.}, \opencite{Manchester_2003}, \opencite{Brown_etal_2003}, \opencite{Kusano_etal_2004}, \opencite{ZhangY_etal_2008}), {\it etc}. 
It is often found that the magnetic free energy accumulated in an active region (AR) exceeds the energy 
required for an eruption. A well studied case is AR~11158 based on the SDO/HMI vector magnetograms~(see, {\it e.g.}, \opencite{Schrijver_etal_2011},
\opencite{Sun_etal_2012}, \opencite{WangS_etal_2012}, \opencite{Vemareddy_etal_2012a}). 
With the aid of a non-linear force-free field (NLFFF) extrapolation method~\cite{Wiegelmann_etal_2012}, \inlinecite{Sun_etal_2012}
investigated the evolution of the magnetic field and its energy in the AR from 12\,--17\, February 2011.
It was found that the magnetic energy continuously increased with the free energy well above $10^{32}$ 
erg. The only X-class flare during the period of interest consumed only a small fraction of the 
accumulated free energy (of the order of 10--20\%). Thus, a pivotal and much unclear issue is what the effective triggers 
of the free energy release are. 

It is now acknowledged that there are generally two kinds of triggering mechanisms. The first is a non-ideal process, associated with magnetic reconnection. The tether-cutting model~\cite{Moore_etal_2001} and magnetic breakout model~\cite{Antiochos_etal_1999} are both of this type. 
The other is loss of equilibrium, an ideal process, due to some instabilities, {\it e.g.} the kink instability~(see, {\it e.g.}, \opencite{Hood_Priest_1979}, \citeyear{Hood_Priest_1980}), torus instability~(see, {\it e.g.}, \opencite{Torok_etal_2004}, \opencite{Kliem_Torok_2006}, \opencite{Fan_Gibson_2007}) and catastrophe~\cite{Forbes_Priest_1995,Lin_Forbes_2000,HuY_2001}. 

\begin{figure*}[tb]
\centering
\includegraphics[width=\hsize]{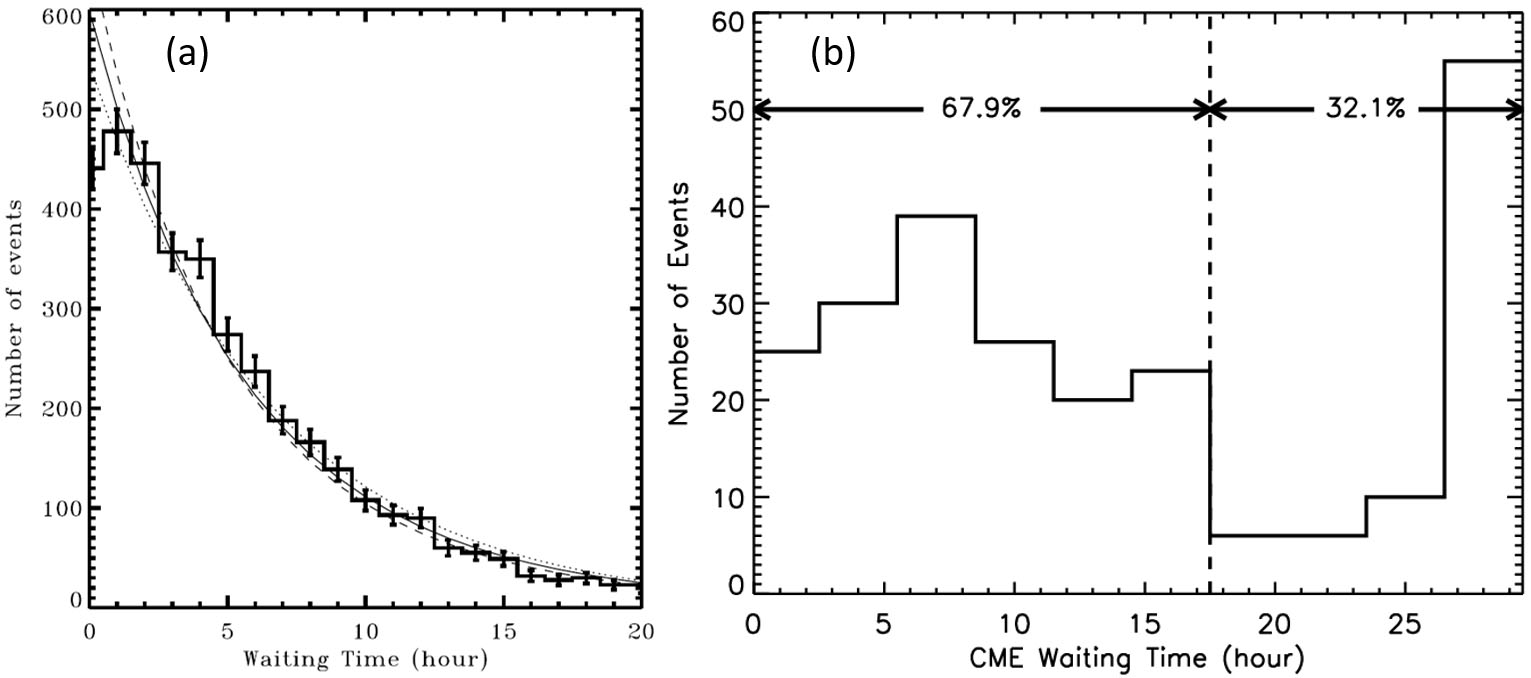}
\caption{(a) Adapted from Moon {\it et al.} (2003b), showing the waiting time distribution of all CMEs during October 1998  -- 
 December 2001. For comparison, a stationary Poisson distribution (dotted line) and two non-stationary Poisson distributions (dashed and solid lines) are plotted. (b) Adapted from Wang {\it et al.} (2013), showing the waiting time distribution of quasi-homologous CMEs originating from all the CME-rich super ARs in Solar Cycle 23. The two panels are reproduced by permission of the American Astronomical Society (AAS).}
\label{fg_sec2_waiting_time_distritions}
\end{figure*}

CMEs are large-scale structures that may involve multiple magnetic flux systems, but trigger points usually start
locally. A question, whether or not the CME occurrence is random, is naturally raised. In other words, can a CME trigger another one and, if yes, how? 
A way to test the degree of inter-dependence of CMEs is through a statistical approach. 
An early attempt to examine the independence of CMEs was done by \inlinecite{Moon_etal_2003}, who
considered 3817 CMEs listed in the LASCO CME catalog~\cite{Yashiro_etal_2004} during 1999--2001.
They generated the waiting time distribution of these CMEs in terms of their first 
appearance in the field of view of LASCO/C2, and found that it is very close to an exponential 
distribution (Figure~\ref{fg_sec2_waiting_time_distritions}a) and can be well explained by a time-dependent 
Poisson random process. A similar distribution can also be found in solar flares~\cite{Wheatland_2000}.
These results imply that interrelated CMEs only constitute, at most, a small fraction of the whole population of CMEs.

On the other hand, modern observations have shown numerous evidence that some CMEs do not occur independently from 
each other. 
Such interrelations can be also found in other explosive phenomena, such as flares,
filament eruptions, {\it etc.}, which are generally referred to as ``sympathetic'' eruptions~(see, {\it e.g.}, \opencite{Richardson_1951}, \opencite{Fritzova-Svestkova_etal_1976}, 
 \opencite{Pearce_Harrison_1990}, \opencite{Biesecker_Thompson_2000},  \opencite{WangH_etal_2001},  \opencite{Moon_etal_2002},  \opencite{Schrijver_Title_2011},  \opencite{Jiang_etal_2011},  \opencite{ShenY_etal_2012},  \opencite{Yang_etal_2012}, \opencite{WangR:2016}). In general, sympathetic CMEs refer to those originating from different regions, 
but almost simultaneously~\cite{Moon_etal_2003}, whereas the eruptions occurring successively from the same 
region in a relatively short interval (several hours), having similar morphology and similar associated phenomena, 
are referred to as homologous CMEs~\cite{Zhang_Wang_2002} or generally called ``quasi-homologous'' CMEs regardless of their 
morphology and associations~(see, {\it e.g.}, \opencite{Chen_etal_2011}, \opencite{Wang_etal_2013}). The two kinds of interrelated 
CMEs are potential candidates for CME-CME interactions, and such interactions may begin during the initiation and last all the way
to the IP space. Thus, it becomes of particular interest to determine under which circumstances CMEs are triggered successively. 

\subsection{Homologous CMEs}\label{sec:homologous}

The possibility that the Sun produces homologous eruptions based on their similar visual aspects and origins was raised at the beginning of space-based coronal observations using ground-based coronagraphs as well as the coronagraph onboard the {\it Orbiting Solar Observatory 7} (OSO-7: \opencite{Hansen:1974}).
Although the waiting times of all CMEs are approximately exponentially distributed, a quite different
distribution can be found if one considers only the waiting times for CMEs originating from the same ARs. 
Based on the source locations of all the CMEs during 1997--1998~\cite{Wang_etal_2011},
\inlinecite{Chen_etal_2011} investigated 15 CME-rich ARs which produced more than 80 quasi-homologous CMEs, and analyzed the waiting times between CMEs from the same AR. It was found that the distribution has two components, clearly separated at around 15 hours. The component within 15 hours follows a Gaussian-like distribution with the peak at around 8 hours 
and it is thought to represent physically related events.
The CMEs in the other component are most likely to be independent. \inlinecite{Wang_etal_2013} extended
the sample to all the CME-rich super ARs in Solar Cycle 23 covering 281 CMEs, and found a similar distribution of
the waiting times of the CMEs (Figure~\ref{fg_sec2_waiting_time_distritions}b). The only difference is that the separation 
time of the two components slightly increases from 15 hours to 18 hours and the peak of the Gaussian-like component
decreases to around 7 hours. In this way, we may refine the definition of quasi-homologous CMEs as the 
successive CMEs originating from the same AR with a separation less than $\sim$ 15--18 hours. 

This finding raises two subsequent questions: how are the quasi-homologous CMEs physically related and what causes the second CME? The Gaussian-like component of the waiting time distribution suggests that either (1)
the magnetic free energy and/or helicity accumulate and reach a threshold on a pace of about 7 hours on average, 
or (2) the timescale of the growth of the instability of a loop system triggered by the preceding CME
is about 7 hours. The former mechanism is applicable to the quasi-homologous CMEs originating from the same polarity inversion lines (PILs),
whereas the latter is for those from the different parts of a PIL or neighboring PILs even though they are in the same AR.
This picture is worthy of further validations with observations.

One widely studied case is the homologous CMEs occurring from AR~9236 on 24\,--25\,November 2000~(see, {\it e.g.}, \opencite{Nitta_Hudson_2001}, \opencite{Zhang_Wang_2002}, \opencite{
Moon_etal_2003a}). In a 60-hour 
interval, a total of six halo CMEs associated with five X-class and one M-class flares originated from 
the AR. By combining {\it Yohkoh} X-ray data and SOHO/MDI magnetograms, \inlinecite{Nitta_Hudson_2001} 
showed that all of the associated flares occurred around the leading spot of the AR. The first four
flares successively originated from the western part of the spot with the emission intensity decreasing.
The intensity of the last two flares increased but originated from the southern part of the spot.
The hard X-ray footpoints were located in different regions for the first four flares as compared to the last two 
flares, suggesting that the two sets of CMEs might originate from the different PILs.
Since many small polarity pairs emerged into the spot during the period, \inlinecite{Nitta_Hudson_2001} suggested
that the continuously emerging magnetic flux was the cause of the successive CMEs and flares. In more details,
\inlinecite{Zhang_Wang_2002} investigated the magnetic flux emergence around the flaring regions for the first three eruptions. They used time-sequences of the high-resolution MDI magnetograms to follow the evolution of 452 moving magnetic features from their births to deaths, and found that there were three flux peaks in
the temporal evolution, which well corresponded to the occurrence of the eruptions. The calculation of
the magnetic helicity based on the MDI magnetograms also showed that there were significant spikes in
the helicity change rate during the eruptions~\cite{Moon_etal_2003a}. These results match the first aforementioned scenario
that the rebuilding of free energy is probably a key mechanism for the homologous CMEs.
It is noteworthy that the first 
three CMEs in the series traveled with increasing speeds from about 700 to 1000~km\,s$^{-1}$, and 
were followed by another extremely fast CME with a speed of $>2000$~km\,s$^{-1}$ originating from a 
different region~\cite{Nitta_Hudson_2001}. These four successive CMEs 
interacted in interplanetary space and formed a complex structure at 1 AU~(\opencite{Wang:2002}, also see Section~\ref{sec:helio}.4).

\begin{figure*}[tb]
  \centering
  \includegraphics[width=\hsize]{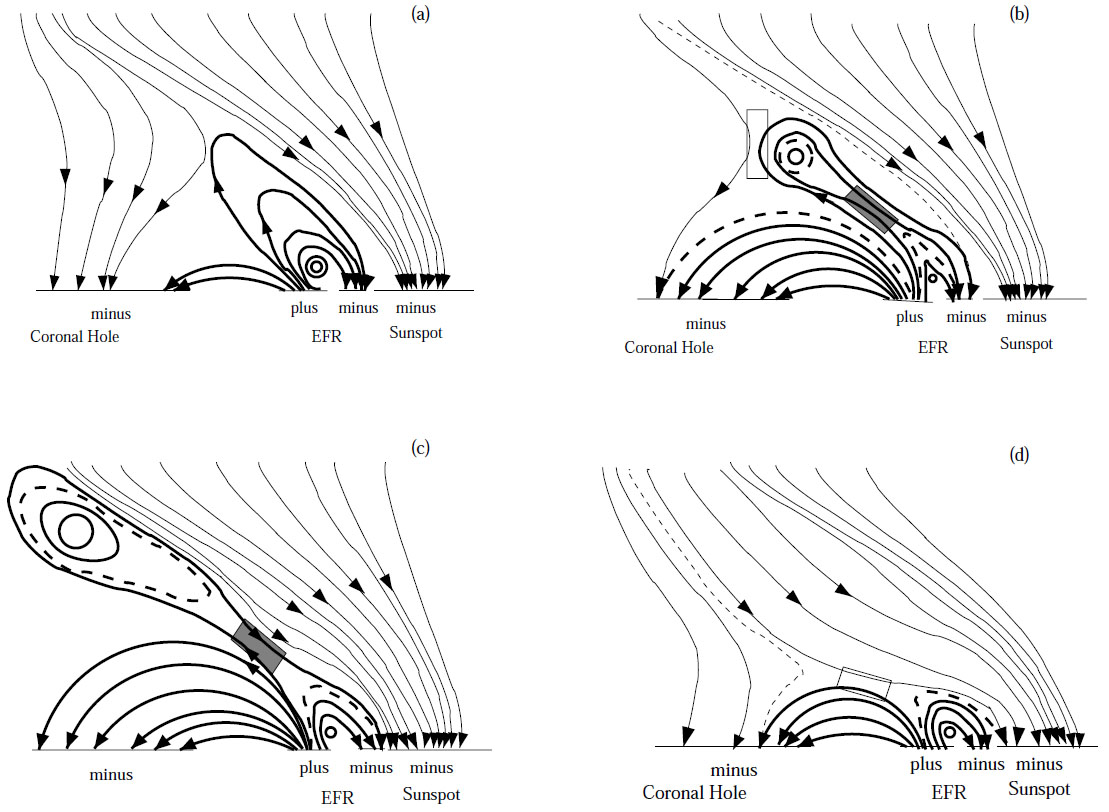}
  \caption{Schematic diagram illustrating the process of continuously emerging fluxes causing homologous CMEs (directly 
adapted from Sterling and Moore, 2001). The rectangles indicate the reconnection regions. Reproduced by permission of the AAS.}
  \label{fg_sec2_homologous_sterling_moore}
\end{figure*}

The process of how the continuously emerging fluxes cause homologous CMEs was previously proposed by \inlinecite{Sterling_Moore_2001} based on the ``breakout'' picture~\cite{Antiochos_etal_1999}. They studied two homologous CME-associated flares 
from AR~8210 on 1\,--\,2 May 1998, and found there were signatures of reconnection between the closed field of the emerging flux and the open field in a neighboring coronal hole. This led to a series of CMEs, as the whole process repeats (see Figure~\ref{fg_sec2_homologous_sterling_moore}). 
Nevertheless, two homologous CMEs reported and studied by \inlinecite{Chandra_etal_2011} seemed to
have different triggering mechanisms. The two CMEs originated from AR~10501 on 20 November 2003, associated 
with homologous H$\alpha$ ribbons. By applying a linear force-free field (LFFF) extrapolation method~\cite{Demoulin_etal_1997},
the authors identified the quasi-separatrix layers in 3D, and compared with the locations of flaring ribbons.
They suggested that the first CME and flare were triggered by the tether-cutting process, which manifested a significant
shear motion and reconnection below the core field, and resulted in a destabilized magnetic configuration for the 
second CME and flare, which were more likely to be initially driven by an instability or a catastrophic process. A similar case was reported by \inlinecite{Cheng:2013}, who studied
two successive CMEs originating on 23 January 2012, and found that the first CME partially removed the overlying
field and triggered the torus instability for the second CME one and half hours later.  These two eruptions have also been studied in details by \inlinecite{Li:2013}, \inlinecite{Joshi_etal_2013} and \inlinecite{Sterling:2014}, and their interplanetary consequences by \inlinecite{Liu:2013}. Another example was the two eruptions separated by about 50 minutes on 7 March 2012 from AR~11429 analyzed  by \inlinecite{WangR:2014}, who studied the magnetic field restructuring and helicity injection changes before and during these two successive eruptions.

Regarding to the time delay between (quasi-)homologous CMEs, an extreme case is that two CMEs originate from one AR at
almost the same time, {\it i.e.} within minutes, the so-called ``twin-CME'' scenario \cite{Li:2012}. One such case was reported by \inlinecite{Shen_etal_2013}, two CMEs launched from AR~11476 within about 2 minutes based on high-resolution and high-cadence observations from SDO. SDO/HMI magnetograms suggest that the CMEs originated from two segments of 
a bent PIL, above which a mature flux rope and a set of sheared arcades were located as revealed by a NLFFF extrapolation.
The twin-CMEs caused the first ground-level enhancement (GLE) event in  Solar
Cycle 24 on 17 May 2012, consistent with the statistical studies that interaction of two CMEs launched in close temporal succession 
favors particle accelerations~(see, {\it e.g.}, \opencite{Li:2012}, \opencite{Ding:2013}). More discussions about the effect of 
interacting CMEs on particle accelerations are continued in Section~\ref{sec:SEP}.

\begin{figure*}[tb]
  \centering
  \includegraphics[width=\hsize]{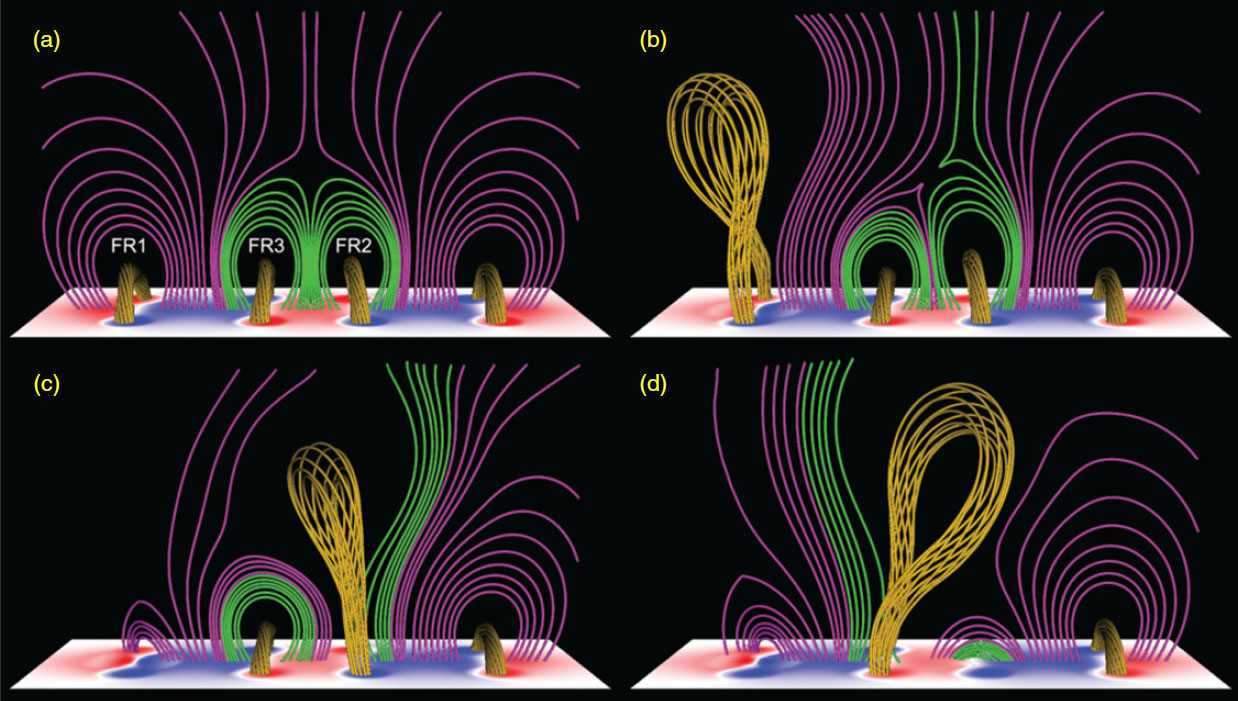}
  \caption{Numerical simulations showing the trigger and initiation of successive CMEs (adapted from T{\"o}r{\"o}k {\it et al.}, 2011). 
The flux ropes, original closed field lines, and open field lines are indicated in yellow, green, and purple colors, respectively. Reproduced by permission of the AAS.}
  \label{fg_sec2_sim_torok}
\end{figure*}

Such successive CMEs from the same AR can be studied in numerical simulations
either by supplying free energy into the system through flux emergence~\cite{MacTaggart_Hood_2009, Chatterjee:2013}, 
through continuous shear 
motions~\cite{DeVore_Antiochos_2008,Soenen_etal_2009}, or by the perturbation of previously neighboring eruptions~\cite{Torok_etal_2011,Bemporad_etal_2012}.
The latter may be treated as a kind of CME-CME interaction during the initiation phase. 
The simulation by \inlinecite{Torok_etal_2011} was established on a set of zero-$\beta$ compressible ideal MHD equations~\cite{Torok_Kliem_2003} in which
four flux ropes~\cite{Titov_Demoulin_1999} were inserted with two of them under a pseudo-streamer and the other two
placed on each side of the pseudo-streamer. After the triggering of the eruption of one flux rope next to the pseudo-streamer, 
the whole simulated system becomes unstable (Figure~\ref{fg_sec2_sim_torok}). The first erupted flux rope expands as it rises and causes breakout reconnection above one of the flux ropes beneath the pseudo-streamer, which leads to the second eruption.
As a consequence, a vertical current sheet forms beneath the second erupted flux rope, and reconnection occurs, which results in a third eruption. Both the second and third eruptions are due to the weakening of the constraints of the overlying fields, suggesting that the torus instability plays a pivotal role in the successive eruptions. The second and third eruptions come from the same pseudo-streamer, and therefore match the picture of quasi-homologous CMEs from the same AR but different PILs. The typical timescale of the torus instability, which leads to the third eruption, is of interest, as the statistical analysis suggests about 
7 hours. However, studies on this point are rare. In addition, the above simulation results might be also applicable to 
sympathetic CMEs, which are discussed next.

\subsection{Sympathetic CMEs}\label{sec:sympathetic}

\begin{figure*}[tb]
  \centering
  \includegraphics[width=\hsize]{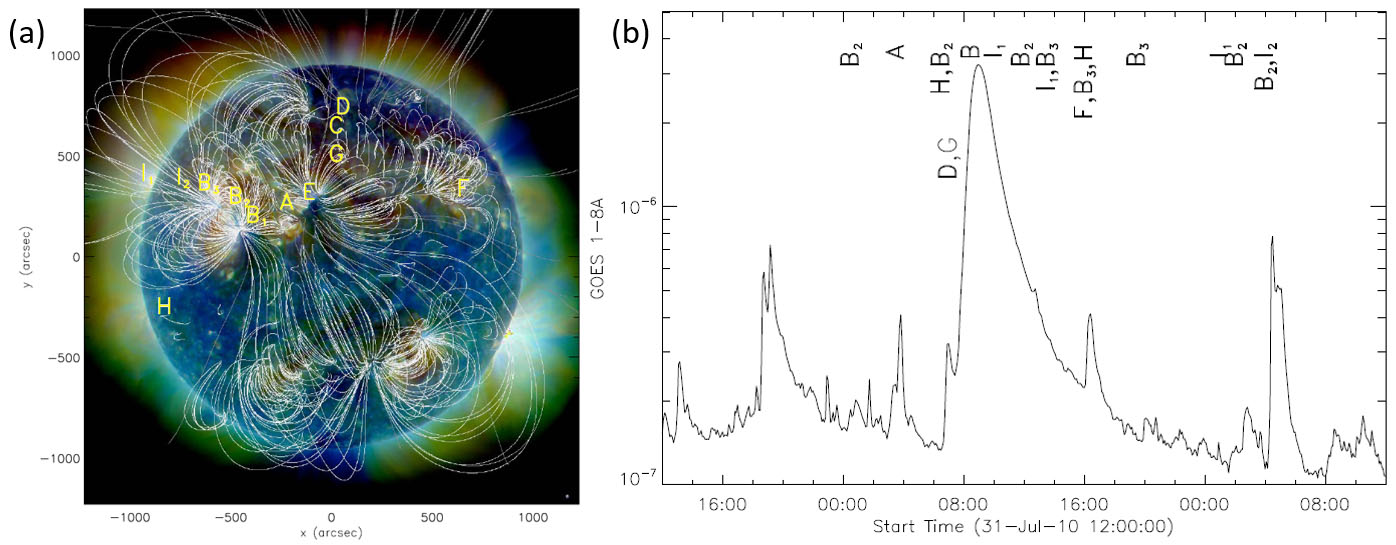}
  \caption{a) Three-color composite EUV image combined from SDO/AIA 211\AA, 193\AA\, and 171\AA\ channels on 1 August 2010. 
Coronal magnetic field lines extrapolated using a potential field source surface (PFSS) model is superimposed showing the magnetic connections among different regions.
Letters note the locations of the eruptive events during 1\,--\,2 August 2010. (b) GOES 1-8\AA\ light curve with the same letters
denoted. Adapted from Schrijver and Title (2011).}
  \label{fg_sec2_sympathetic_schrijver}
\end{figure*}

As defined earlier, sympathetic CMEs originate almost simultaneously but from spatially separated regions, and one eruption contributes to the triggering of another one. \inlinecite{Lyons:1999} mentioned the possibility for ``one CME [to] activate the onset of another'', whereas \inlinecite{Moon_etal_2003} are the first to use specifically the term ``sympathetic CMEs''. 
Defining the term ``simultaneously'' quantitatively is a complex problem. In most studies,
it refers to temporal separation between the eruptions of less than several hours. Thus, in this aspect, sympathetic CMEs are similar to those quasi-homologous CMEs originating from different PILs in the same AR. The key question for sympathetic CMEs is how distant magnetic systems 
connect and interact with each other in such a short interval. 
The study by \inlinecite{Simnett_Hudson_1997} showed that the CME occurring on 23 February 1997 erupted from the north-east limb of the Sun and quickly merged with a previously much larger event, which was associated with a loop system connecting the northern region to the southern region (another
example can be found in Figure~\ref{fg_sec2_sympathetic_schrijver}). Such
transequatorial loops are not rare. A statistical study based on {\it Yohkoh} data from October 1991 to December 1998
showed that one third of all ARs present transequatorial loops~\cite{Pevtsov_2000}, suggesting that ARs can be magnetically  
connected even though they are located on the opposite hemispheres of the Sun (see also \opencite{Webb:1997}).

\inlinecite{WangH_etal_2001} presented a case of the connection between two M-class sympathetic flares from two different ARs (referred to as  inter-AR interaction). The two flares 
apart by about 1.5 hours originated from AR~8869 and 8872 on 17 February 2000. Both were associated with a filament. During the progress of the first flare, the associated filament disappeared
and a loop structure connecting the two flaring regions became visible in H$\alpha$ images. Along the path of the loop,
a surge starting from one end of the erupted filament quickly excited a set of disturbances propagating toward 
the other AR, which was followed by the second flare and the second filament disappearance. The speed of the disturbances
was estimated as about 80~km\,s$^{-1}$, close to the local Alfv{\'e}n speed. Another similar interaction between
two eruptions was presented in \inlinecite{JiangY_etal_2008}, in which a transequatorial jet disturbed inter-AR loops and led to
an eruption of the inter-AR loops. The jet and the loop eruptions drove two CMEs separated by less than 2 hours. 
Combining multi-wavelength observations including the higher-resolution data from SDO, \inlinecite{Joshi_etal_2016} recently
described sympathetic eruptions in two adjacent ARs on 17 November 2013. A scenario of a series of chain reconnections 
was proposed for these eruptions with the aid of a NLFFF extrapolation.

Such connections or interactions are not limited to adjacent ARs. Thanks to the stereoscopic observations provided by 
STEREO twin spacecraft as well as SOHO and SDO near the Earth, the global connections among flares and CMEs originating
from different regions can be explored. A well-studied series of events are the interrelated eruptive events during 1\,--\,2 August 
2010~(see, {\it e.g.}, \opencite{Schrijver_Title_2011}, \opencite{Harrison:2012}, \opencite{Liu:2012}). The study by \inlinecite{Schrijver_Title_2011} focused on
the near-synchronous long-distance interactions between magnetic domains.
They identified more than ten events including flares, filament eruptions and CMEs. With the aid of a magnetic field extrapolation method based on the potential field assumption, they 
investigated the global topology of the magnetic field and its changes. It was found that all the scattered major 
events were connected via large-scale separators, separatrices and quasi-separatrix layers.  
These results are consistent with the study by \inlinecite{Titov_etal_2012}, who also 
reconstructed the topology of the coronal magnetic field and investigated the connections between the eruptions and 
the pseudo-streamers, separatrices and quasi-separatrix layers. They proposed that reconnections along these separators
triggered by the first eruption probably caused the sequential eruptions. 
The resulting CMEs interacted with each other during their propagation in interplanetary space. A more complete picture 
of this series of events is given in \inlinecite{Harrison:2012} and \inlinecite{Liu:2012}. The long-distance coupling was further studied with more events by \inlinecite{Schrijver_etal_2013}. They argued that there are several distinct pathways for sympathetic
eruptions, {\it e.g.}, waves or propagating perturbations, distortion of or reconnection with the overlying field by distant 
eruptions and other (in)direct magnetic connections. 

The simulations by \inlinecite{Torok_etal_2011}, mentioned in Section~\ref{sec:homologous}, 
reproduced the successive CMEs from the regions beneath and beside a pseudo-streamer (Figure~\ref{fg_sec2_sim_torok}), which is not only applicable to the eruption of quasi-homologous 
CMEs from one AR but also to the possible long-distance coupling between different ARs. In their simulations,
the breakout reconnection and weakening of overlying fields due to the neighboring eruptions are responsible 
for sequential eruptions. The same process was reproduced in the 2.5D MHD simulations by \inlinecite{Lynch_Edmondson_2013}.
With a full 3D MHD code under the Space Weather Modeling Framework~\cite{Toth_etal_2012,Holst_etal_2014}, \inlinecite{Jin_etal_2016} numerically
studied the long-distance magnetic impacts of CMEs. The coronal environment on 15 February 2011 was established and
a CME was initiated by inserting a flux rope of \inlinecite{Gibson_Low_1998} analytical solution into AR 11158. The impacts
of the CME on eight ARs, five filament channels, and two quiet Sun regions were evaluated by the decay index, defined as 
$-\frac{d\log B(h)}{d\log h}$, where $B$ is the magnetic field and $h$ is the height above the solar surface, and other 
impact factors. 
They found that the impact gets weaker at longer distances and/or for stronger magnetic structures, and suggested that there were two different types of the impacts. The first is the direct impact due to the CME expansion and 
the induced reconnection, which may efficiently weaken the overlying field. It is limited spatially to the CME expansion domain. The second is indirect impact outside the
CME expansion domain, where the impact of the CME is propagated through waves during both the 
eruption and the post-eruption phases, and the overlying field may be weakened especially when 
the global magnetic field relaxes to a steady state during the post-eruption phase. 

Although the mechanisms of long-distance coupling have been extensively studied and well documented, it is still unclear 
under which circumstances a CME may successfully take off. That is to say, not all of the regions impacted by a CME do
launch a sequential CME. 
The same issue holds for (quasi-)homologous eruptions.

 \section{Effects of Successive CMEs on Particle Acceleration}\label{sec:SEP}

\subsection{Successive CMEs and Solar Energetic Particle Events}

Solar energetic particles (SEPs) are known to be accelerated in association with two main phenomena: solar flares and CMEs. Historically, SEPs have been divided into impulsive events of shorter duration, most often associated with solar flares but not always \cite{Kahler:2001c} and gradual events, most often associated with CME-driven shocks \cite{Cane:1986,Reames:1999,Cliver:2004}. There are typically significant differences between SEPs accelerated through these two mechanisms, including the duration, elemental abundances, spectra, {\it etc.}\ \cite{Mason:1999,Desai:2003,Desai:2006,Tylka:2005}. 
In the past two decades, with remote observations of CMEs and {\it in situ} measurements of SEPs, it has become well established that the largest gradual SEP events are associated with fast and wide CMEs \cite{kahler92,Reames:1990b,Zank:2000}. Large CME shock fronts are ideal accelerators for charged particles and therefore, SEPs can occasionally reach energies up to several GeVs. SEPs together with cosmic rays play an important role in Space Weather (see, {\it e.g.}, \opencite{usoskin13}). 

The magnetic field configuration is crucial in order to determine whether accelerated particles might be detected or not. For Earth-affecting SEP events, the particles are thought to be injected onto field lines located in the western hemisphere of the Sun, accounting for Sun-Earth connecting magnetic structures due to the Parker spiral shape of the IP magnetic field (see, {\it e.g.}, \opencite{klein08}, \opencite{schwenn06}). Therefore, fast CMEs originating from the western hemisphere of the Sun are more likely to be magnetically connected to Earth; and hence, fast and wide, western-limb CMEs are the most common cause of large gradual SEP events \cite{Cane:1988,Gopalswamy:2004}. There are also large SEP events observed with clear sources from the eastern solar hemisphere. From STEREO observations, with widely separated spacecraft, it is recognized that SEPs are indeed widespread phenomena (see, {\it e.g.}, \opencite{Dresing:2012}). However, a simple look at SEP and CME statistics reveals that not all fast, wide, and western CMEs are associated with large SEP events \cite{Ding:2013}. 

Different scenarios of acceleration processes for electrons and ions have been discussed (see, {\it e.g.}, \opencite{kliem03}). Among others, coronal waves, CME lateral expansion as well as CME-CME interaction are possible candidates. Studies on Moreton and EUV waves are still unresolved and cannot fully rule out coronal waves as SEP driving agent (against: \opencite{bothmer97}, \opencite{krucker99}, \opencite{miteva14}; pro: \opencite{malandraki09}, \opencite{rouillard12}). CME-CME interaction itself might play a minor role in the SEP production, but a preceding CME might have a significant effect in terms of preconditioning. This idea originated from a statistical study by \inlinecite{Gopalswamy:2002} which showed that the presence of a previous CME within 12 hours of a wide and fast CME greatly increases the probability that this second, fast CME is SEP-rich \cite{Gopalswamy:2002,Gopalswamy:2004}. The reverse relation was also found: SEP-rich CMEs are about three times more likely than average to be preceded by another eruption \cite{Kahler:2005}. In another study of 57 large SEP events that had intensities $>$ 10 pfu (particle flux units, 1 pfu = 1 proton~cm$^{-2}$~s$^{-1}$~sr$^{-1}$) at $>$ 10 MeV/nuc, \inlinecite{Gopalswamy:2004} showed that there exists a strong correlation between high particle intensities and the presence of preceding CMEs within 24 hours of the main SEP-accelerating CME. As the acceleration of SEPs is believed to happen within the first 10~$R_\odot$ (and most likely within the first 4-5~$R_\odot$), this timeline makes it less probable that direct shock-shock interaction is responsible for the observed larger probability of SEP events \cite{Richardson:2003,Kahler:2003}. While important, these studies are not enough to determine the physical causes of these statistical relations. Hence, the role of interacting CMEs and their relation to large SEP events still leaves many questions open. 

\begin{figure}    
   \centerline{\includegraphics[width=0.95\textwidth,clip=]{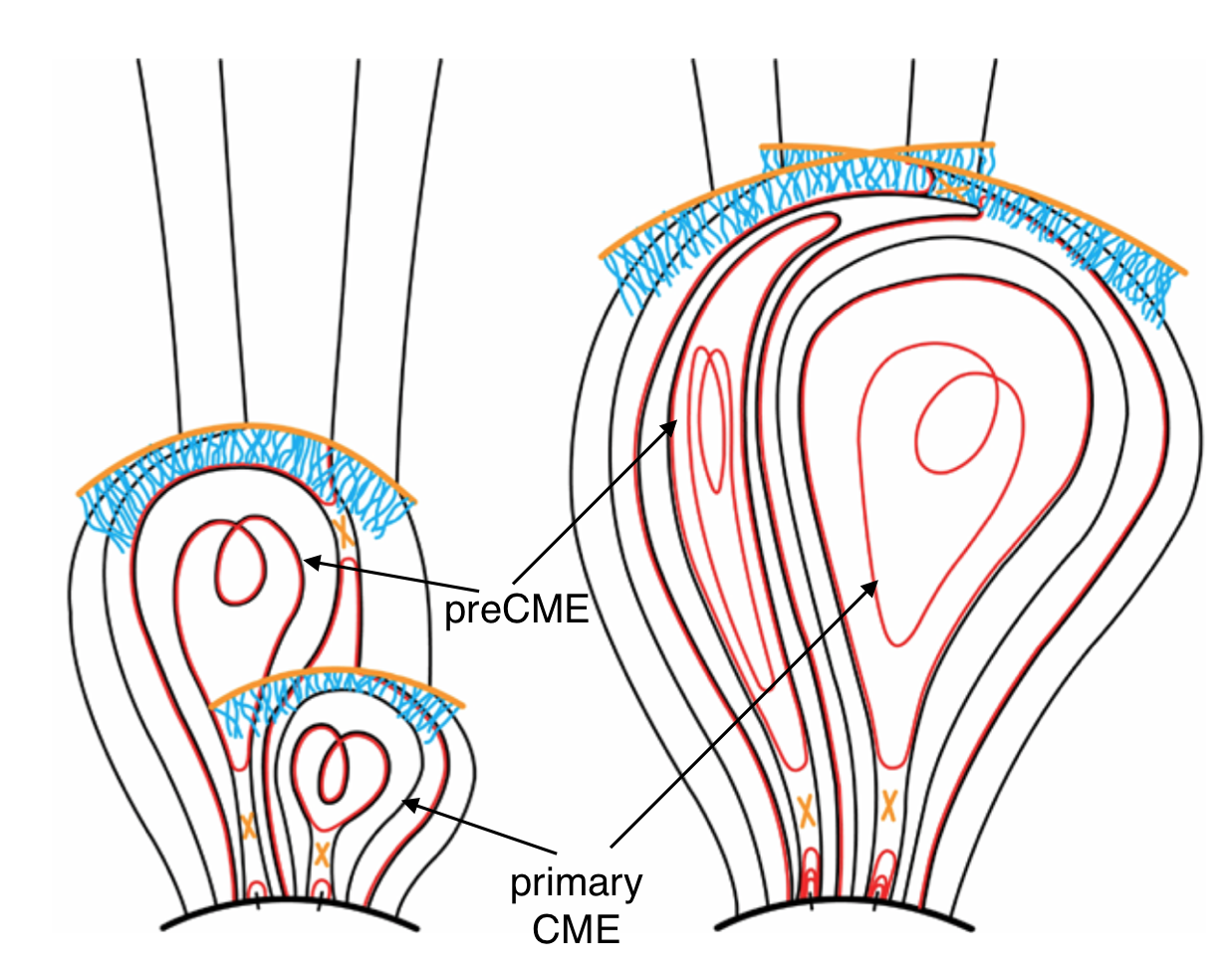}              }
\caption{The twin-CME scenario first outlined by Li {\it et al.} (2012) and adapted by Kahler and Vourlidas (2014). Left: the preCME drives a turbulent shock region (blue shaded area). The SEP-producing CME (primary CME) is launched close to the preCME but later in time. The magnetically accessible (interchange reconnection, marked by orange crosses) turbulent shock region in the preCME acts as amplifier for particles accelerated by the shock of the primary CME. Right: the more developed phase of the preCME-CME interaction, where the primary CME shock has crossed the reconnection region.}
   \label{fig0}
   \end{figure}

The preconditioning of the ambient environment close to the Sun has important effects on SEP production. 1) Preceding CMEs (pre-CMEs) not only can provide enhanced seed population, but also lead to a stronger turbulence at the second shock, therefore, increasing the maximum energy of the particles (this is referred to as the twin-CME scenario, as proposed by \inlinecite{Li:2005} and further developed in \inlinecite{Li:2012}; see Figure~\ref{fig0} and Section~\ref{sec:homologous}). \inlinecite{Ding:2013} and \inlinecite{Ding:2014b} tested the twin-CME scenario against all large SEP events and fast CMEs with speeds $>$ 900~km\,s$^{-1}$ from the western hemisphere in Solar Cycle 23. They suggested that a reasonable choice of the time threshold for separating a single CME and a twin-CME is 13 hours. Using this time delay, they found that 60\% twin-CMEs lead to large SEPs while only 21\% single CMEs lead to large SEPs. Furthermore, all large SEP events with a peak intensity larger than 100 pfu at $>$ 10 MeV/nuc recorded by the {\it Geostationary Operational Environmental Satellite} (GOES) are twin-CMEs. Note that twin-CMEs may or may not be associated with direct interaction between the CMEs themselves.  2) The change of the nature of closed and open magnetic field lines in the vicinity of an AR may result in a different shock angle. \inlinecite{Tylka:2005} and \inlinecite{Sokolov:2006}, among others, have shown that shock geometry can have a large influence on the SEP flux and intensity. 3) The presence of closed field lines within a CME might trap particles accelerated by a subsequent CME and, hence, decrease the flux of high-energy particles at Earth \cite{Kahler:2003}, or  increase their maximum energy \cite{Gopalswamy:2004}. 4) On longer timescales, the presence of a CME in the heliosphere might dramatically modify the Sun-Earth magnetic connectivity, the length and solar footprints of the field lines connected to Earth. This is clearly visible when a SEP event occurs while an interplanetary CME (ICME) passes over Earth \cite{Kallenrode:2001b,Ruffolo:2006}. This type of configuration usually results in delaying SEPs, but it might also significantly change the Sun-Earth connectivity \cite{Richardson:1991}. \inlinecite{Masson:2013} investigated how flare-accelerated SEPs may reach open field lines through magnetic reconnection during a CME-associated flare. Similar processes need to occur during CME-CME interaction for accelerated particles to be measured at Earth. 

Although the association of preceding CMEs with enhanced SEP intensity is a robust observation, alternative explanations to the twin-CME scenario exist. In a recent work, \inlinecite{Kahler:2014}, making use of an extensive SEP list from \inlinecite{Kahler:2013}, found a relation between the 2 MeV proton background intensities and an increase in the SEP event intensities and the occurrence rates of preceding CMEs. They suggested that preceding CMEs may be an observational signature of enhanced SEP intensities but are not physically coupled with them. This is in contradiction to the events studied in \inlinecite{Gopalswamy:2004} and \inlinecite{Ding:2013} for which no association of larger SEP events with $>$ 2 MeV backgrounds is found. We note that most of the CME related studies are based on the LASCO CME catalogue \cite{Yashiro_etal_2004}, which contains measurements of CME kinematics and hence, energies at heights too far away (beyond 10~R$R_\odot$) to be directly compared with particle energies. Therefore the importance of the background effect remains unclear, as well as whether 2~MeV particles are the right energy level to study ``seed'' particles for SEPs or not.

\subsection{Radio Signatures of CME-CME Interaction}

\begin{figure}    
   \centerline{\includegraphics[width=0.95\textwidth,clip=]{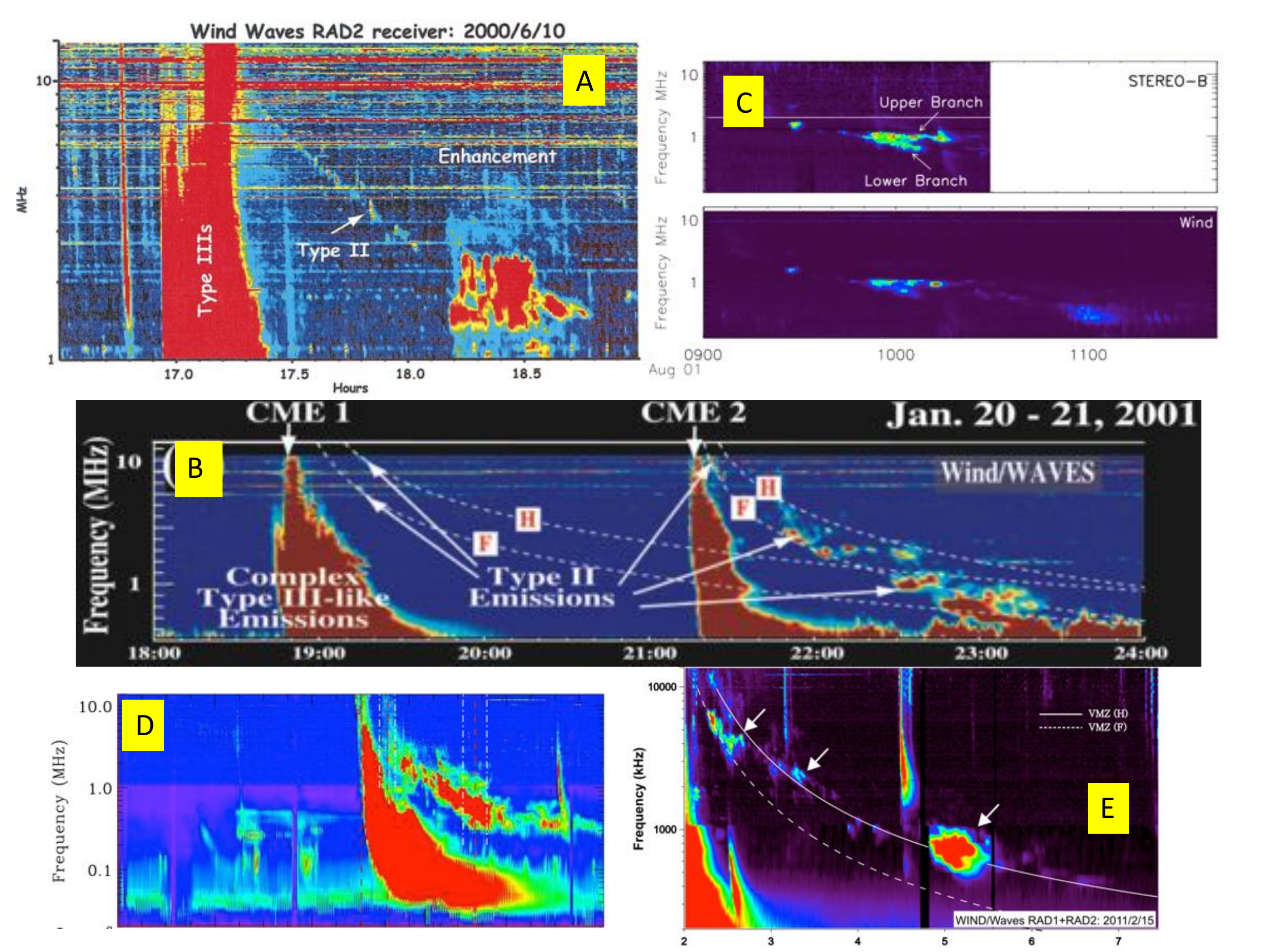}}
\caption{Selection of observations for enhanced type II radio bursts associated to CME-CME interaction events (identified by the solar event naming convention - solar object locator SOL): a) Gopalswamy {\it et al.}, 2001 (SOL-2000-06-10); b) Reiner {\it et al.}, 2003 (SOL-2001-01-20); c) Martinez-Oliveros {\it et al.}, 2012 (SOL-2011-08-01); d) Ding {\it et al.}, 2014 (SOL-2013-05-22); e) Temmer {\it et al.}, 2014 (SOL-2011-02-15).}
   \label{fig1}
   \end{figure}

Closely related to CMEs and SEP production processes, and most probably more closely related to CME-CME interaction events, is the observation of enhanced radio emission for CME-CME events. \inlinecite{Gopalswamy:2001}  first reported about radio signatures in the long-wavelength range, that occurred as intense continuum-like radio emission following an interplanetary type II burst. They linked the timing of the enhancement in the radio emission to the overtaking of a slow CME by a faster one. As shown in Figure~\ref{fig1}, enhanced radio signatures as consequence of CME-CME interaction are in fact frequently reported (see, {\it e.g.}, \opencite{Reiner:2003}, \opencite{hillaris11}, \opencite{martinez-oliveros12}, \opencite{Ding:2014}, \opencite{Temmer:2014}). We note that the description of such a scenario is intimately connected to the 3D geometry and propagation direction of two CMEs. While many of the studies are in agreement that the CME interaction is the cause of the radio enhancement, the interpretation is not straightforward. We review this process step by step. 

Type II radio bursts give information on the propagation and density behavior of the CME associated shock component \cite{mann95}. As a consequence of interacting CMEs, a continuum-like enhancement of decametric to hectometric (dm to hm) type II radio emission may be interpreted as observational signature of the transit of the shock front of the fast CME through the core of the slow CME. This presumes that the upstream compression due to the passage of a CME enhances the particle density and, therefore, decreases the background Alfv{\'e}n velocity, which would result in a stronger shock \cite{Gopalswamy:2004,Kahler:2005,Li:2005}. However, the collision not only increases the electron density due to compression but also the magnetic field. In fact, the Alfv{\'e}n speed is expected to be higher inside a CME that would actually lead to a reduction of the shock Mach number \cite{Kahler:2003,klein06}. Even with higher coronal density, this would make the overtaking shock weaker and less likely to occur. Numerical simulations actually show that in CME-CME interaction events, large variations in density, Alfv{\'e}n speed, and magnetic field can be expected within the preceding CME \cite{Lugaz:2005b}. In this respect, we note that a reduced Alfv{\'e}n speed within the structures would reduce the efficiency of reconnection processes and CME ``cannibalism'' might not  work efficiently. This is confirmed by \textit{in situ} measurements from CME-CME interaction events showing rather intact separate flux ropes for the CME-CME interaction events (see, {\it e.g.}, \opencite{martinez-oliveros12}). Nevertheless, the merging process is taking place as shown, among others, by \inlinecite{Maricic:2014} where possible reconnection outbursts from \textit{in situ} data at 1~AU are observed for the CME-CME interaction event series from 13\,--\,15 February 2011. However, the time span needed to merge two flux ropes completely might be too short and might only be observed beyond distances of 1~AU -- more details about CME-CME interaction processes are found in Section~\ref{sec:helio}.

\inlinecite{Ding:2014} and \inlinecite{Temmer:2014} are two of the few examples using stereoscopic observations with the Graduated Cylindrical Shell (GCS) model of \inlinecite{Thernisien:2011} to determine the real direction and heights of two successively erupting CMEs rather than plane-of-sky heights and projected directions. Using this approach, \inlinecite{Ding:2014} found that the start time of type II radio emissions coincided with the interaction between the front of the second CME and the trailing edge of the first CME, interaction which occurred around 6~$R_\odot$, also close to the distance of peak SEP acceleration. This is not supported by \inlinecite{Temmer:2014} who concluded that the timing for the enhanced type II bursts did not match the time of interaction for the CMEs, but they could be related to a kind of shock-streamer interaction \cite{shen13}. In the event under study, the flanks of the following CME might interact with the field which was opened and compressed by the preceding CME. Another scenario describing the occurrence of continuum-like radio emissions might be reconnection processes of the poloidal field components between the interacting CMEs \cite{Gopalswamy:2004}.  In fact, enhanced type II radio signatures may be the signatures of several different types of interactions.

Information on the magnetic field topology involved in the process of CME-CME interaction might be given by radio type III bursts. Radio type III bursts are generated by energetic electron beams guided along quasi-open magnetic field lines. Due to a sudden change in the magnetic topology, type N radio bursts show in addition a drift in the opposite direction (classical interpretation: magnetic mirror effect). As CMEs manifest themselves as sudden change in the, generally outward directed, IP magnetic field and electron density distribution, \inlinecite{demoulin07} conclude that decametric type N radio bursts are most likely not caused by mirroring effects but due to geometry effects as consequence of the magnetic restructuring in CME-CME interaction events. \inlinecite{hillaris11} report on peculiar type III radio bursts due to accelerated electrons that might be disrupted by the turbulence near the front of a preceding CME (see also \opencite{reiner99}). Results from \inlinecite{Temmer:2014} showed that the observed type II enhancements, which were associated with type III bursts, stopped at frequencies related to the downstream region of the extrapolated type II burst, as if it would be a barrier for particles entering the magnetic structure \cite{macdowall89}. 

In this respect, there have been several attempts to explore magnetic connectivity to interplanetary observers. Some of them have used realistic MHD simulations combined with a simple particle source input at the inner boundary in the inner heliosphere and ballistic particle propagation \cite{Luhmann:2010}, while others employ an idealized shock surface and Parker spiral, together with physics-based transport \cite{Aran:2007,Rodriguez-Gasen:2014}. \inlinecite{Masson:2012} investigated the interplanetary magnetic field configurations based on observations during ten GLE events, and concluded that particle arrival times were significantly later than what would be expected under a Parker spiral field, illustrating how the magnetic connectivity to a given observer cannot be assumed to be static. It may be modified before and during the eruption: by other structures between the Sun and the Earth, such as other CMEs, solar wind streams and corotating interaction regions (CIRs), or by reconnection occurring close to the solar surface. Recently, \inlinecite{Kahler:2016} tested the appropriateness of the Parker's spiral approximation for SEP studies using the Wang-Sheeley-Arge (WSA: \opencite{Wang:1990}, \opencite{Arge:2000}) model, and reached similar conclusions. One limitation of these studies is that none of them includes a magnetic ejecta driving a shock wave which is initiated in the low corona, {\it i.e.}\,where particle acceleration is known to occur. 

The significant drawback in these observational studies comes from the limitation of currently available data, which may only reveal the consequences of CME-CME interaction but not the interaction process itself. A more direct insight into the CME-CME interaction process and related plasma and magnetic field parameters could be gained from {\it in situ} data. However, most of CME events collide far from where plasma and magnetic field parameters are actually monitored. An exception is the 30 September 2012 event, which revealed interaction between two CMEs close to 1AU (probably started interacting $\sim$0.8 AU) as shown in studies by \inlinecite{Liu:2014b} and \inlinecite{Mishra:2015b}. The {\it in situ} instruments aboard {\it Solar Orbiter} and {\it Solar Probe+} (to be launched in 2018), which will travel at close distances to the Sun, will be of great interest and will give a great complementary view on the CME-CME interaction processes. 
  
\section{The Interaction of CMEs in the Inner Heliosphere}\label{sec:helio}

Direct observations of CME-CME interaction became possible in the mid-1990s with the larger field-of-view of LASCO/C3 (up to 32~R$_\odot \sim$ 0.15~AU) which yielded the first reported white-light observations of CME-CME interaction \cite{Gopalswamy:2001}. Although the interaction of successive CMEs at distances beyond LASCO/C3 field-of-view can often be deduced from their white-light time-distance track or from radio emissions \cite{Reiner:2001}, only few articles focused on the analysis of direct interaction following this first report \cite{Reiner:2003}. In the meantime, there had been a resurgence of interest regarding CME-CME interaction based on the analysis of {\it in situ} measurements near L1 \cite{Burlaga:2002,Burlaga:2003,Wang:2002,Wang:2003,Wang:2003a}. 

Reported observations became relatively routine with the development of heliospheric imaging, first with SMEI starting in 2003 and, second with the heliospheric imagers (HIs) onboard STEREO starting in 2007 (see Figure~7 for some examples). Although a number of SMEI observations focused on series of CMEs \cite{Bisi:2008,Jackson:2008}, their analyses did not dwell on the physical processes occurring during CME-CME interaction. However, one of the very first CMEs observed remotely by STEREO was in fact a series of two interacting CMEs \cite{Harrison:2009,Lugaz:2008b,Lugaz:2009b,Lugaz:2009c,Odstrcil:2009,Webb:2009}. During the period from the first remote detection in 2001 to routine remote observations in the late 2000s, numerical simulations have been used to fill the gap between the upper corona and the near-Earth space. Early simulations include the work by \inlinecite{Wu:2002}, \inlinecite{Odstrcil:2003} and \inlinecite{Schmidt:2004}. In the past decade, the combination of these three approaches (remote observations, {\it in situ} measurements and numerical simulations) has resulted in a much deeper understanding of the physical processes occurring during CME-CME interaction.

\subsection{Changes in the CME Properties}

One of the essential aspects of CME-CME interaction is the change in CME properties, such as their speed, size, expansion rate, {\it etc}. This may directly affect  space weather forecasting, as not only the CME speed and direction may change (modifying the hit/miss probability and the expected arrival time) but also its internal magnetic field (modifying the expected geomagnetic responses). In addition, understanding how CME-CME interaction changes CME properties can deepen our understanding of the internal structure of CMEs.
Many studies have investigated the change in the speed of CMEs due to their interaction, both through remote observations and numerical simulations. Some studies have focused on the nature of the collision, in terms of restitution coefficient and inelastic {\it vs.}\ elastic {\it vs.}\ super-elastic collision \cite{CShen:2012,Mishra:2015a,Mishra:2015b,Colaninno:2015,Mishra:2016}. Different natures of collision seem to be possible (see for example the review by \opencite{FShen:2016b}). Particularly, if CME-CME collision can be super-elastic, it raises the questions of which circumstances yield an increase of the total kinetic energy, and what is the source of the kinetic energy gain. 
However, CMEs are large-scale magnetized plasma structures propagating in the solar wind, and therefore, the ``collision'' of CMEs is a much more complex process than the classic collision of ordinary objects. There are many factors causing the complexity: 1) depending on the speed of the CMEs, the interaction between two CMEs may involve zero, one or two CME-driven shocks, some of which may dissipate during the interaction, 2) the interaction should take at least one Alfv{\'e}n crossing time of a CME. With a typical CME size at 0.5~AU of 20--25~R$_\odot$, and a typical Alfv{\'e}n speed inside a CME of 200-500~km~s$^{-1}$, the interaction should take 8--24 hours, 3) the CME speed can change significantly, even at large distances from the Sun due to their interaction with the solar wind \cite{Temmer:2011,WuCC:2016}, 4) CME-CME interaction is inherently a three-dimensional process, and the changes in kinematics may differ greatly depending on the CME part that is considered \cite{Temmer:2014}. Using numerical simulations, it is somewhat possible to control some of these effects, for example by performing simulations with or without interaction but identical CME properties, and by knowing the velocity field in the entire 3D domain \cite{FShen:2013,FShen:2016}. This has revealed that the momentum exchange with the ambient solar wind during CME-CME interaction may be neglected in some cases.

\begin{figure*}[tb]
   \centerline{\includegraphics[width=0.98\textwidth,clip=]{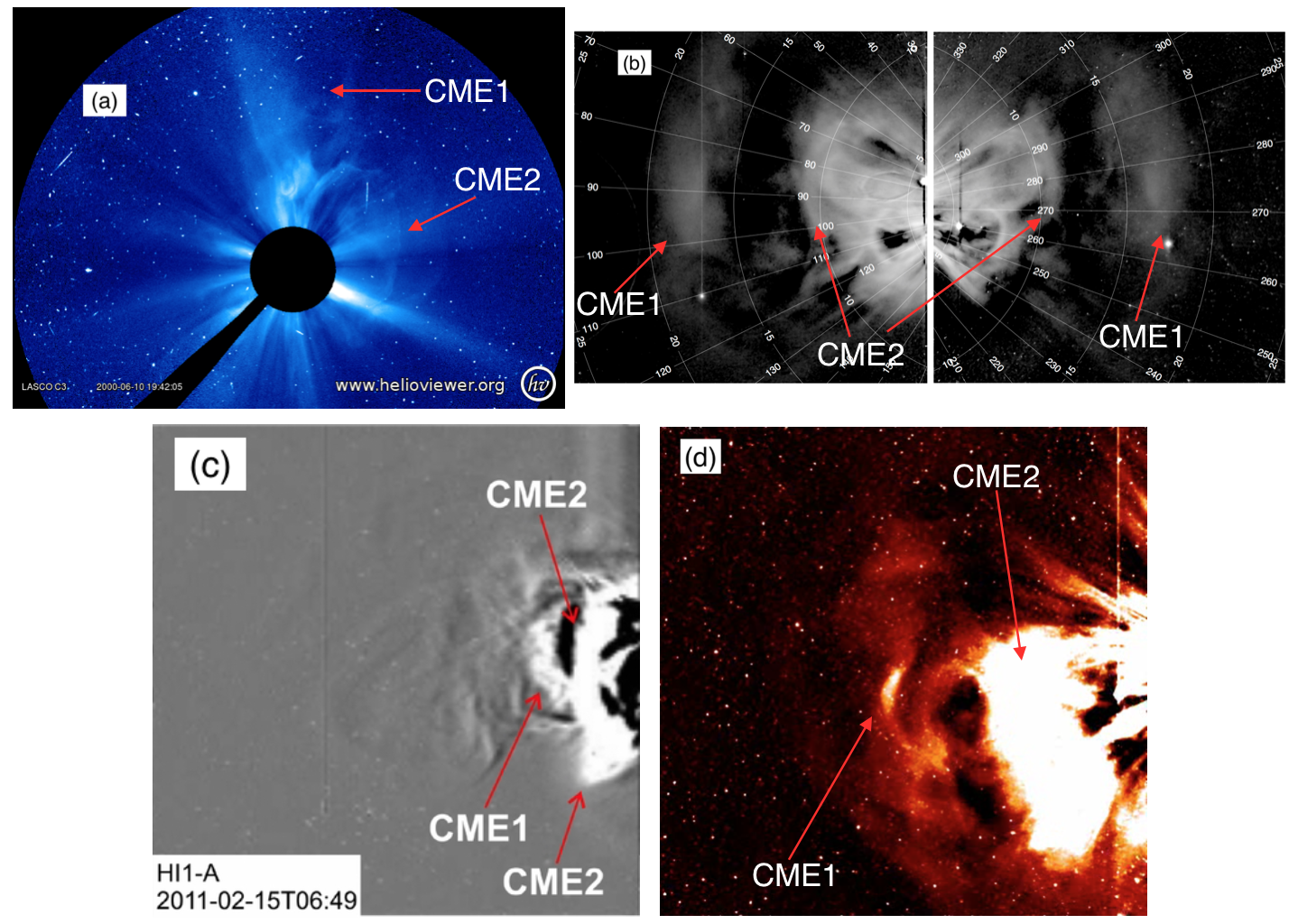}}
\label{fig:CME_HI}
\caption{Observations of CME-CME interaction in LASCO/C3 and STEREO/HI1 fields-of-view. (a) shows the two CMEs from the initial report of CME-CME interaction by Gopalswamy {\it et al.} (2001). (b) shows base-difference images on 25 May 2010 at 01:29UT (left: HI1A, right: HI1B) corresponding to the event studied in Lugaz {\it et al.} (2012). (c) is a running-difference HI1A image of the event of 15 February 2011 studied in Temmer {\it et al.} (2014). (d) is a base difference image of the HI1A image of the 10 November 2012 event studied by Mishra, Srivastava and Chakrabarty (2015). (c) is reproduced by permission of the AAS.}
\end{figure*}

As CME-CME interaction involves a faster, second CME overtaking a slower, leading CME, the end result is to homogenize the speed, as was noted from {\it in situ} measurements in \inlinecite{Burlaga:2002}, \inlinecite{Farrugia:2004}, and through simulations by \inlinecite{Schmidt:2004} and \inlinecite{Lugaz:2005b}, among others, and this occurs independently of the relative speed of the two CMEs. One main issue is to understand what determines the final speed of the complex ejecta which was formed through the CME-CME interaction. In an early work, \inlinecite{Wang:2005} found that, in the absence of CME-driven shock waves, the final speed is determined by that of the slower ejecta, whereas \inlinecite{Schmidt:2004} and \inlinecite{Lugaz:2005b} found that when the CMEs drive shocks, the final speed is primarily determined by that of the faster ejecta, as the shock's propagation through the first magnetic ejecta accelerates it to a speed similar to that of the second ejecta (see Figure~8). 

Most recent works have combined remote observations and numerical simulations. It now appears relatively clear that the final speeds depend on the relative masses of the CMEs, as well as their approaching speed \cite{FShen:2016}, and, hence, on their relative kinetic energy. \inlinecite{Poedts:2003}, based on 2.5D simulations, noted that the acceleration of the first CME increases as the mass of the second CME increases and that an apparent acceleration of the first CME is in fact due to a slower than expected deceleration. Making the situation more complex are the changes in the CME expansion during the propagation as well as the fact that remote observations can be used to determine the velocity of the dense structures, but not really that of the low-density magnetic ejecta. As discussed in \inlinecite{Lugaz:2005b} and further in \inlinecite{Lugaz:2009c}, when the trailing shock impacts the leading magnetic ejecta, the dense sheath behind the shock must remain between the two magnetic ejecta, even as the shock propagates through the first ejecta. As HIs observe density structures, the observations may not be able to capture the shock propagating through a low-density ejecta.

\begin{figure*}[tb]
  \centering
   \centerline{\includegraphics[width=0.98\textwidth,clip=]{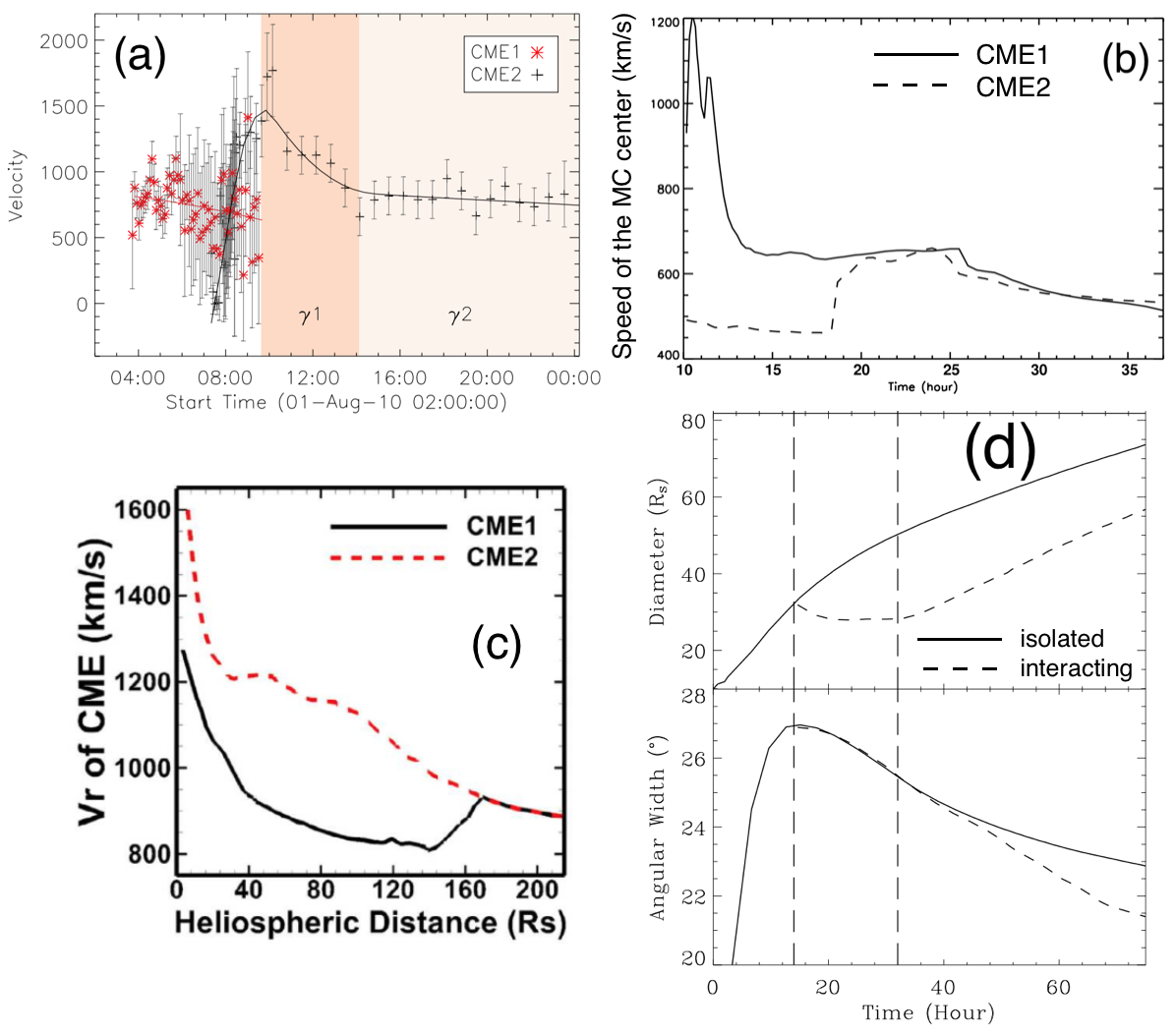}}
\label{fig:CME_speed}
\caption{Changes in the CME properties due to CME-CME interaction. (a) shows the changes in speed associated with the August 2010 events from Temmer {\it et al.} (2012); see also Liu {\it et al.} (2012). (b) and (c) show the change of speeds for simulated CMEs from the work of Lugaz {\it et al.} (2005) and  Shen {\it et al.} (2012b), respectively. (d) shows the change in radial size and angular width for a case with a shock overtaking a CME (dash) {\it vs}.\ an isolated CME (solid) from Xiong {\it et al.} (2006). (a) and (b) are reproduced by permission of AAS. (c) and (d) are reproduced by permission of John Wiley and Sons.}
\end{figure*}

In addition to changes in velocity, CME-CME interaction may result in the deflection of one CME by another \cite{Xiong:2009,Lugaz:2012b,CShen:2012}. Combining these works, it appears that the deflection can reach up to 15$^\circ$ when the two CMEs are initially about 15\,--\,20$^\circ$ apart. Such angular separations are quite frequent between successive CMEs, as it corresponds to a delay of about one day for two CME originating from the same active region (due to solar rotation). This change in direction must be taken into account when deriving the changes in velocity, as done in \inlinecite{CShen:2012} and \inlinecite{Mishra:2016}. 

Next, we discuss the changes in the CME internal properties, such as radial extent, radial expansion speed, and magnetic field strength. Only the radial extent can be reliably derived from remote observations \cite{Savani:2009,Nieves:2012,Lugaz:2012b}. Both numerical simulations and remote observations confirm that the radial extent of the leading CME plateaus during the main phase ({\it i.e}.\ when the speed of both CMEs changes significantly) of interaction \cite{Schmidt:2004,Lugaz:2005b,Xiong:2006,Lugaz:2012b,Lugaz:2013b} and this is typically associated with a ``pancaking'' of the leading CME \cite{Vandas:1997}. It should be noted that it appears nearly impossible for the CME radial extent to decrease, but rather the compression of the back of the leading CME is associated with a slowing down of its radial expansion. In \inlinecite{Lugaz:2005b}, the authors discussed how the shock propagation through the leading CME is the main way in which the expansion slows. It is as yet unclear whether the compression changes for cases with or without shocks. What clearly changes is the resulting expansion of the leading CME after the end of the main interaction phase ({\it i.e}.\ after the shock exited the ejecta). 
In \inlinecite{Xiong:2006}, the authors found that the leading CME overexpands to return to its expected size; as such the compression is only a temporal state. This was confirmed by the statistical study of the magnetic ejecta radial size at different distances \cite{Gulisano:2010}, as well as one study where remote observations indicated compression but, a day after the interaction ended, when the CME impacted Earth, the {\it in situ} measurements indicated a typical CME size \cite{Lugaz:2012b}. In numerical simulations with two magnetic ejecta (see Figure~9), the rate of over-expansion is found to depend on the rate of reconnection between the two ejecta; as such, it depends on the relative orientation of the two magnetic ejecta \cite{Schmidt:2004,Lugaz:2013b}, but also probably on their density. The potential full coalesence of two ejecta into one was discussed in a few studies \cite{Odstrcil:2003,Schmidt:2004,Chatterjee:2013} but has not been investigated in details with realistic reconnection rates. 
        
\subsection{Changes in the Shock Properties}

In addition to changes in the CME properties, the fast forward shocks propagating inside the magnetic ejecta encounter highly varying and unusual upstream conditions, affecting the shock properties. Most of what is known about the changes in shock properties was learnt from numerical simulations; however, there have been many reported detections of shocks propagating inside a magnetic cloud or magnetic ejecta at 1~AU \cite{Wang:2003a,Collier:2007,Richardson:2010b,Lugaz:2015a,Lugaz:2016}. 

\inlinecite{Vandas:1997} noted that a shock propagates faster inside a magnetic cloud due to the enhanced fast magnetosonic speeds inside, which may result in shock-shock merging close to the nose of the magnetic cloud but two distinct shocks in the flanks. \inlinecite{Odstrcil:2003} noted that associated with this acceleration, the density jump becomes smaller.
\inlinecite{Lugaz:2005b} performed an in-depth analysis of the changes in the shock properties, dividing the interaction into four main phases: i) before any physical interaction, when the shock propagates faster than an identical isolated shock due to the smaller density in the solar wind, ii) during the shock propagation inside the magnetic cloud, when the shock speed in a rest frame increases and its compression ratio decreases, confirming the findings of \inlinecite{Odstrcil:2003}, iii) during the shock propagation inside the dense sheath when the shock decelerates, as pointed out by \inlinecite{Vandas:1997}, and iv) the shock-shock merging when, as predicted by MHD theory, a stronger shock forms followed by a contact discontinuity. If the shock is weak or slow enough, it may dissipate as it propagates into the region of higher magnetosonic speed inside the magnetic cloud \cite{Xiong:2006,Lugaz:2007}. High spatial resolution is necessary to resolve weak shocks in MHD simulations, and low resolution may affect the prediction of shock dissipation. The merging or dissipation of shocks was noted by \inlinecite{Farrugia:2004} when Helios measured four shocks at 0.67~AU and ISEE-3 measured only two shocks later on at 1~AU.
Shock-shock interaction was studied by means of 2-D MHD simulations by \inlinecite{Poedts:2003}, where the authors identified a fast forward shock and a contact discontinuity as the result of two fast forward shocks merging. 

\begin{figure*}[tb]
  \centering
  \includegraphics[width=6cm]{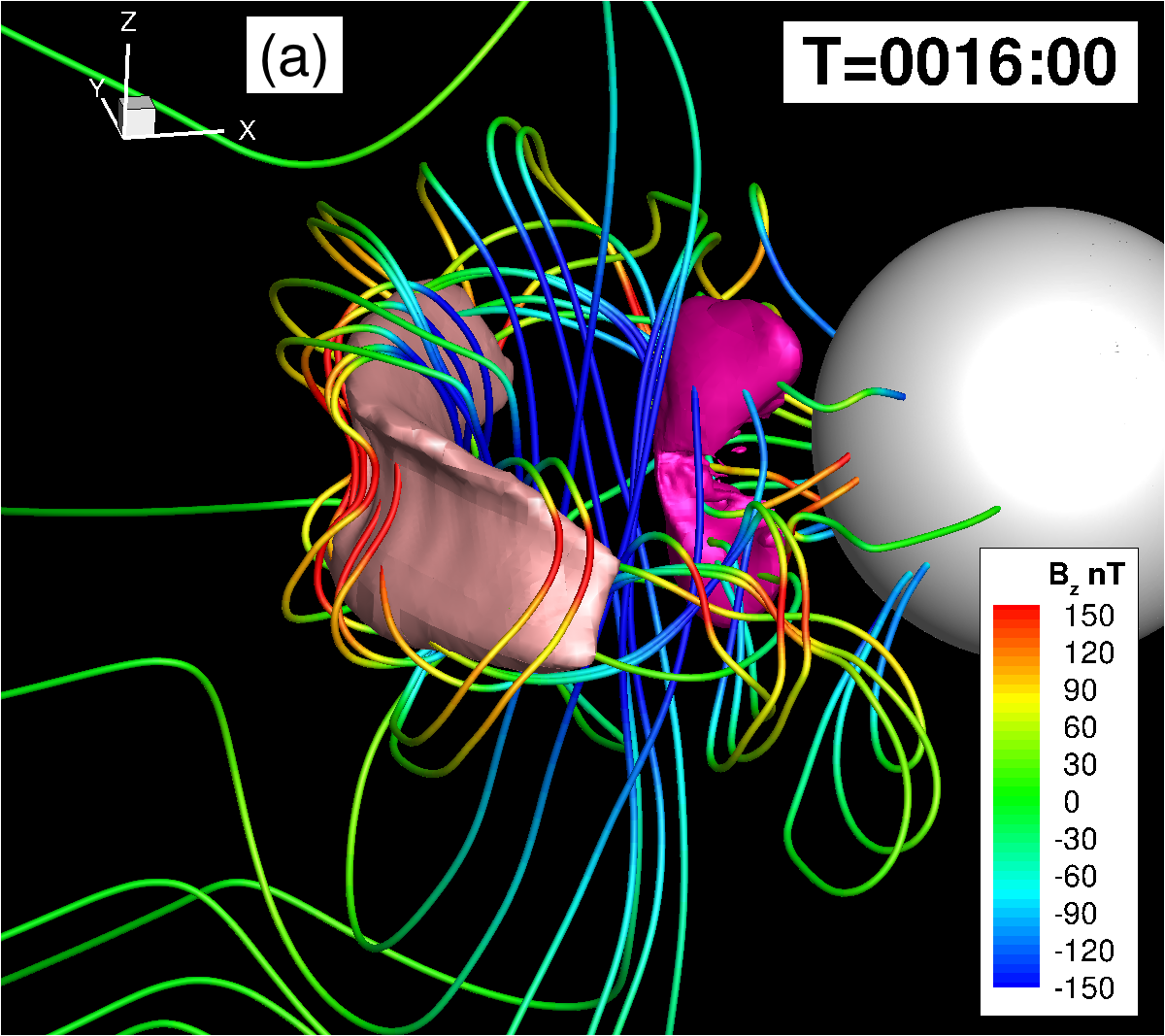}
    \includegraphics[width=6cm]{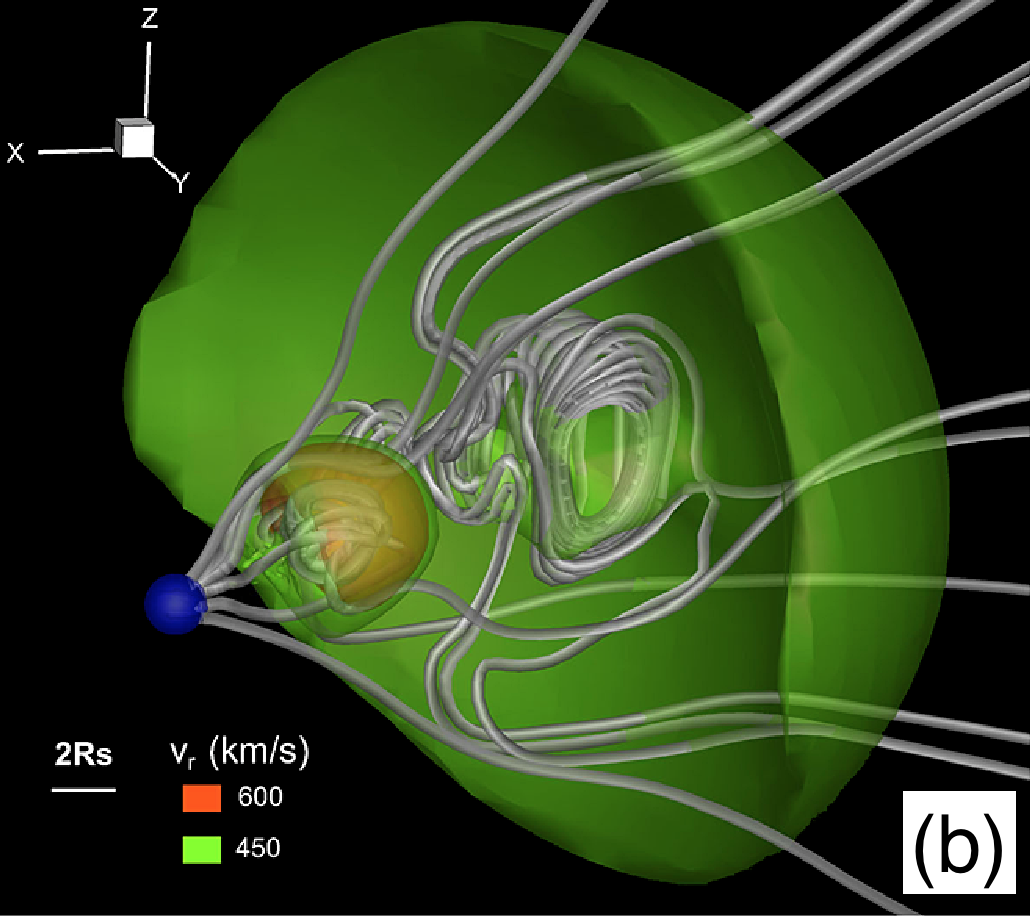}
\label{fig:CME_simu}
\caption{3D MHD simulations of CME-CME interaction. (a) shows a case simulated by Lugaz {\it et al.} (2013) with two CMEs with perpendicular orientations. 3D magnetic field lines are color-coded with the north-south $B_z$ component of the magnetic field. Isosurfaces show regions of east-west $B_y$ component of the magnetic field equal to $\pm$~ 170~nT (pink positive, fuchsia negative). The CME fronts are at about 0.3 and 0.15~AU. (b) shows the simulation of Shen {\it et al.} (2013) with isosurfaces of radial velocity and magnetic field lines in white. (b) is reproduced by permission of John Wiley and Sons.}
\end{figure*}

\subsection{Cases Without Direct Interaction}

In a similar way that the succession but not the interaction {\it per se} of CMEs can affect the resulting flux of SEPs, the succession of CMEs, even without interaction may affect the properties of the second (and subsequent) CMEs. \inlinecite{Lugaz:2005b} performed the simulation of the two CMEs, initiated with the same parameters (initial energy, size, orientation, {\it etc.}) 10 hours apart; the second CME did not decelerate as much as the first one and therefore had a faster speed and a faster shock wave, even before the interaction started. This result was confirmed in studies with different orientations \cite{Lugaz:2013b}. Many of the shortest Sun-to-1~AU transit times of CMEs appear to be associated with a succession of non-interacting CMEs. It has been suggested that it was the case for the Carrington event of 1859 \cite{Cliver:2004b}, the Halloween events of 2003 \cite{Toth:2007}, and the 23 July 2012 event \cite{Liu:2014}, each of which is a case where the propagation lasted less than 20 hours, {\it i.e}.\ the average transit speed was in excess of 2000~km\,s$^{-1}$. Note that only 15 events propagated from Sun to Earth in less than 24 hours in the past 150 years \cite{Gopalswamy:2005}. \inlinecite{Liu:2014} and \inlinecite{Liu:2015} proposed that this succession of non-interacting CMEs may produce a ``perfect storm'' with the most extreme geoeffectiveness. A careful analysis of which situation result in the most geoeffective storms has to be undertaken.
The main reason for this reduced deceleration is that the first CME removed some of the ambient solar wind mass, resulting in less dense and faster flows ahead of the subsequent CME. As such, the second CME experiences less drag and propagates faster \cite{Lugaz:2005b,Liu:2014,Temmer:2015}. 

\subsection{Resulting Structures}

The complex interaction between different shock waves and magnetic ejecta can result in a variety of structures at 1~AU. The ``simplest'' one is a multiple-magnetic cloud (multiple-MC) event \cite{Wang:2002}, in which a single dense sheath precedes two (or more) distinct MCs (or MC-like ejecta). The two MCs are separated by a short period of large plasma $\beta$, corresponding to hot plasma with weaker and more turbulent magnetic field \cite{Wang:2003}, which may be an indication of reconnection between the ejecta (see Figure~10a). Typically, both MCs have a uniform speed profile, {\it i.e}.\ they propagate approximately with the same speed. 
The prototypical example of a multiple-MC event is the 31 March\,--\,1 April 2001 multiple-MC event \cite{Wang:2003,Berdichevsky:2003,Farrugia:2006}. Such structures have been successfully reproduced in simulations \cite{Wang:2005,Lugaz:2005b,Xiong:2007,FShen:2011}. These simulations reveal that the dense sheath ahead of the two MCs may be the result of the merging of two shock waves. In this case, it is expected that the sheath may be composed of a leading hot part (the sheath of the new merged shock) followed by a denser and cooler section (material which has been compressed twice, see \opencite{Lugaz:2005b}). The extremely dense sheath preceding the March 2001 event may be related to a shock-shock merging. It is also possible that the shock driven by the overtaking CME dissipates as it propagates inside the first MC, which would also result in a single sheath preceding two MCs.

Multiple-MC events correspond to cases when the individual MCs can be distinguished, although the uniform speed and the single sheath indicate that they interacted. When multiple ejecta cannot be distinguished, the resulting structure is typically referred to as a complex ejecta or compound stream \cite{Burlaga:2003}. These structures often have a decreasing speed profile, typical of a single event but with complex magnetic fields and a duration of several days (see Figure~10c). Such complex streams may be caused by a number of factors, including 1) interaction close to the Sun, resulting in quasi-cannibalism \cite{Gopalswamy:2001}, 2) the relative orientation of the successive ejecta favorable for reconnection \cite{Lugaz:2013b}, or 3) interaction between more than two CMEs \cite{Lugaz:2007}. 
Some events, for examples that of 26\,--\,28 November 2000, which involved between three and six successive CMEs, have been analyzed as multiple-MC event \cite{Wang:2002} or complex ejecta \cite{Burlaga:2002}. Even if individual ejecta can be distinguished, they are of short duration (for this event between 3 and 8 hours) and the magnetic field is not smooth. In the simulation of \inlinecite{Lugaz:2007}, it was found that the complex interaction of three successive CMEs and the associated compression resulted in a period of enhanced magnetic field and higher speed at 1~AU but without individual ejecta being identifiable. In this sense, complex ejecta at 1~AU are similar to merged interaction regions often measured in the outer heliosphere, corresponding to the merging of many successive CMEs \cite{Burlaga:1997,leRoux:1999}. It is also possible that the interaction between a fast and massive CME and a slow and small CME may result in cannibalism, whereas interaction of CMEs with similar size and energy results in a multiple-MC event. Numerical simulations have focused primarily on the interaction of CMEs of comparable energies and sizes, but a more complete investigation of the effect of different initial sizes and CME energies are required, building up on the work of \inlinecite{Poedts:2003}.

It has also been proposed that seemingly isolated but long-duration events (events that last 36 hours or more at 1~AU) may be associated with the interaction of successive CMEs of nearly perpendicular orientation (see Figure~10b). \inlinecite{Dasso:2009} performed the analysis of the 15 May 2005 CME, including {\it in situ} measurements, radio emissions as well as remote observations (H$\alpha$, EUV, magnetogram and coronagraphic). They concluded that this large event, which lasted close to 2 days at 1~AU was likely to be associated with two non-merging MCs, of nearly perpendicular orientation. The simulations of \inlinecite{Lugaz:2013b} included a case in which two CMEs were initiated with near-perpendicular orientation, as well as two CMEs with the same initial orientation. In the latter case, the authors found that a multiple-MC event was the resulting structure; in the former case, the resulting structure was a long-duration transient having many of the characteristics of a single ejecta. \inlinecite{Lugaz:2014} compared the result at 1~AU of this simulation with the 19\,--\,22 March 2001 CME, another 48-hour period of smooth and slowly rotating magnetic field, monotonically decreasing speed and lower than expected temperature. In both the simulation and data, the second part of the event was characterized by nearly unidirectional magnetic field. The difference between complex ejecta and this type of transient lies in the smoothness of the magnetic field. Both the events studied by \inlinecite{Dasso:2009} and \inlinecite{Lugaz:2014} have been characterized as a single, isolated CME, but their size (twice larger than a typical MC) and the variation of the magnetic field make it unlikely.

\begin{figure*}[tb]
  \centering
  \includegraphics[width=11.5cm]{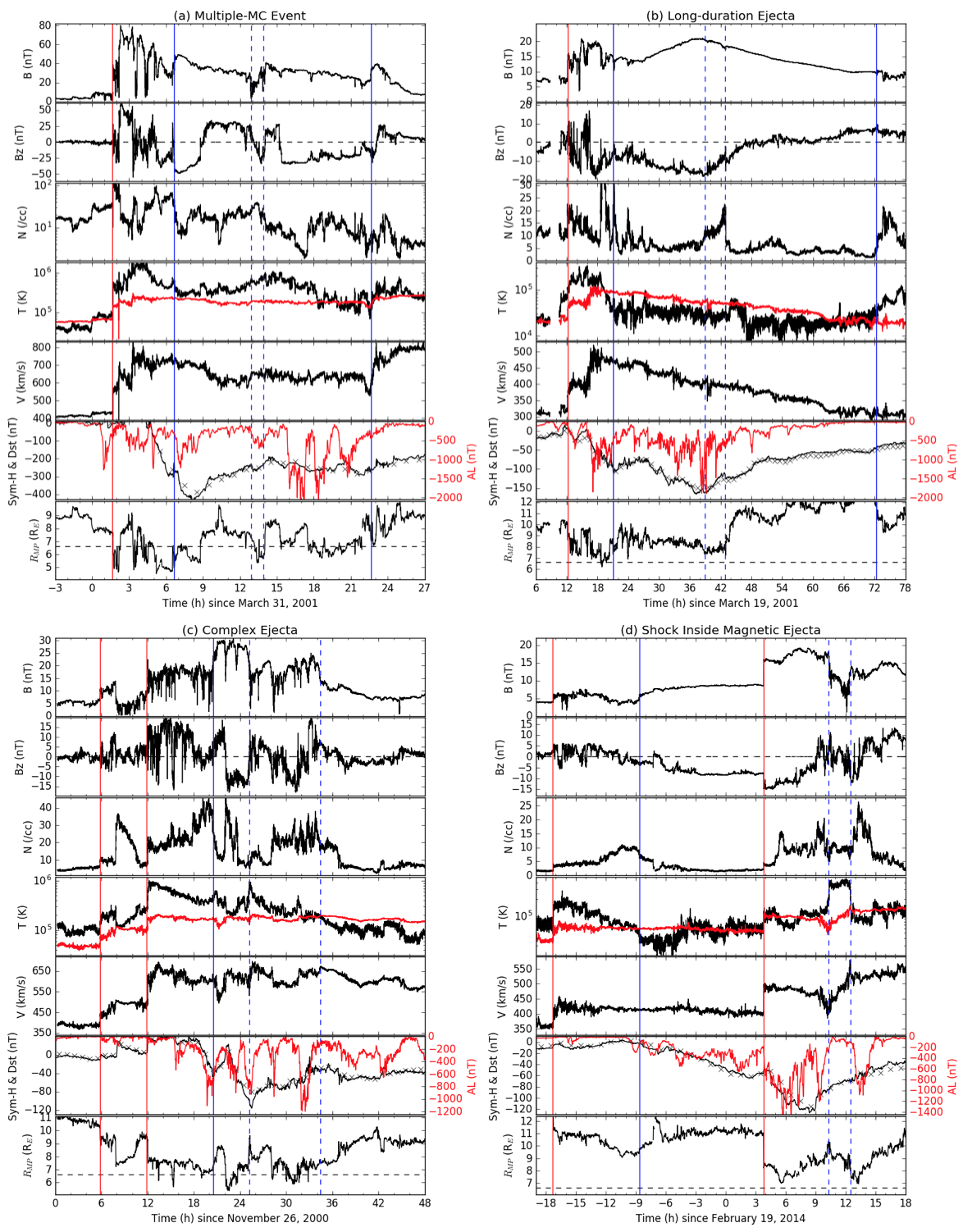}
\label{fig:Insitu}
\caption{{\it In situ} measurements of CME-CME interaction. The panels show the magnetic field strength, $B_z$ component in Geocentric Solar Magnetospheric (GSM) coordinates, proton density, temperature (expected temperature in red), velocity, Sym-H index (Dst with crosses, AL in red), and dayside magnetopause minimum location following Shue {\it et al.} (1998), from top to bottom. Shocks are marked with red lines, and CME boundaries with blue lines (dashed for internal boundaries).}
\end{figure*}

These three cases (multiple-MC, complex ejecta and long-duration event) correspond to full interaction, in the sense that the resulting structure at 1~AU is propagating with a single speed profile (typically monotonically decreasing). The main examples of partial, ongoing interaction are associated with the propagation of a fast forward shock wave inside a preceding ejecta \cite{Wang:2003a,Collier:2007,Lugaz:2015a}. There is a clear difference between the part of the first CME which has been accelerated by the overtaking shock as compared to its front which is still in ``pristine'' conditions (see Figure~10d). In some cases, the back of the first CME is in the process of merging with the front of the second CME \cite{Liu:2014b}, {\it i.e}.\ a complex ejecta or a long-duration event is in the process of forming.
In the study of \inlinecite{Lugaz:2015a}, the authors identified 49 such shocks propagating within a previous magnetic ejecta between 1997 and 2006. Most such shocks occur towards the back of the ejecta, and shocks tend to be slower as they get closer to the CME front. This can be interpreted as an indication that a number of shocks dissipate inside a CME before exiting it. The two main reasons are that CMEs tend to be expanding and have a decreasing speed profile and that the peak Alfv{\'e}n speed typically occurs close to the center of the magnetic ejecta. The latter reason means that shocks become weaker as they approach the center of the ejecta. The former reason implies that shocks propagate into higher and higher upstream speeds as they move from the back to the front of the CME. \inlinecite{Lugaz:2015a} reported cases when the speed at the front of the first CME exceeds the speed of the overtaking shock, {\it i.e}.\ because of the CME expansion, the shock cannot overtake the front of the CME. 

These four different structures, often observed at 1~AU, represent four different ways for CME-CME interaction to affect our geospace in a way which differ from the typical interaction of a CME with Earth's magnetosphere. We give some details on the geo-effectiveness of these structures in the following section.

\section{The Geoeffectiveness of Interacting CMEs}\label{sec:geo-effect}

Compared to isolated ejecta, the geomagnetic disturbances brought about by interacting CMEs need to account for the changes in parameters resulting from their interaction.
These typically, though not always, enhance its geoeffectivess.
Individual CMEs  and their subset magnetic clouds \cite{Burlaga:1981} are known to be major sources of strong magnetospheric disturbances \cite{Gosling:1991}. This is mainly because they often contain a slowly-varying negative north-south ($B_z$) magnetic field component which can reach extreme values (see, {\it e.g.}, \opencite{Farrugia:1993}, \opencite{Tsurutani:1988}). 
The passage of ejecta at 1 AU takes typically about one day, so these disturbances can last for many hours. By contrast, passage of interacting ejecta as well as complex ejecta can take $\sim$3 days at 1 AU \cite{Burlaga:2002,Xie:2006}, so that the magnetosphere is under strong solar wind forcing for a much longer time. Below we show an example of a  storm where the Dst index, which monitors the strength of the ring current, remained below $-200$~nT for about 21 hours. Thus from the point of view of space weather, CME-CME interactions are key players.

Clearly, to assess the overall impact of ejecta interactions on space weather, it is crucial to see how frequent they occur at 1~AU. Some studies have addressed this issue. Examining the causes of major geomagnetic storms in the 10-year period
1996-2005,  \inlinecite{Zhang:2007} showed that at least 27\% of the intense storms (over 88 storms from 1997 to 2005) were due to multiple CMEs. Using a different approach,
\inlinecite{Farrugia:2006b} used the epsilon parameter from \inlinecite{Akasofu:1981} to define so-called ``large events''.  They used this parameter to estimate the energy extracted by the magnetosphere from the solar wind and the powering of the magnetosphere by the solar wind. In the period 1995-2003 they found six out of the 16 largest events  ($\sim$37\%) 
involved CMEs interacting with each other forming
complex ejecta.  One may conclude that CME-CME interactions are important drivers of extreme space weather.  Indeed, some recent studies using multi-spacecraft observations at different radial distances and wide azimuthal separations have been  undertaken illustrating how these interactions play a leading role. A case in point is the interplanetary and geomagnetic consequences of multiple solar eruptions which took place 1 August 2010 \cite{Moestl:2012,Temmer:2012,Liu:2012}.

CME-CME interactions are more frequent during solar maximum conditions when the number of CMEs erupting from the Sun may reach half a dozen {\it per} day. As an illustration, Figure~11 shows {\it in situ} measurements by {\it Wind} of a 31-day period during the maximum phase of Solar Cycle 23. From top to bottom, the figure shows the 
proton density, bulk speed, temperature, the eleven ICMEs identified in this
period (from \opencite{Richardson:2010}), the $B_z$ component of the 
magnetic field (in GSM coordinates), the total field, the proton $\beta$, the 
storm-time Dst index and the planetary Kp index. The red trace in the third panel from the top shows
the expected proton temperature for normal solar wind expansion \cite{Lopez:1987}.
It can be seen that the measured temperature is often well below this value. This is an often-used 
indicator of ejecta material in space (following the initial report by \opencite{Gosling:1973}, used to select CMEs by \opencite{Richardson:1993}, \citeyear{Richardson:1995}). Complementary to this are the episodic high magnetic field strengths. The sawtooth appearance of the temporal profile of bulk speed indicates a series of radially expanding transients.

\begin{figure*}[tb]
  \centering
  \includegraphics[width=11cm]{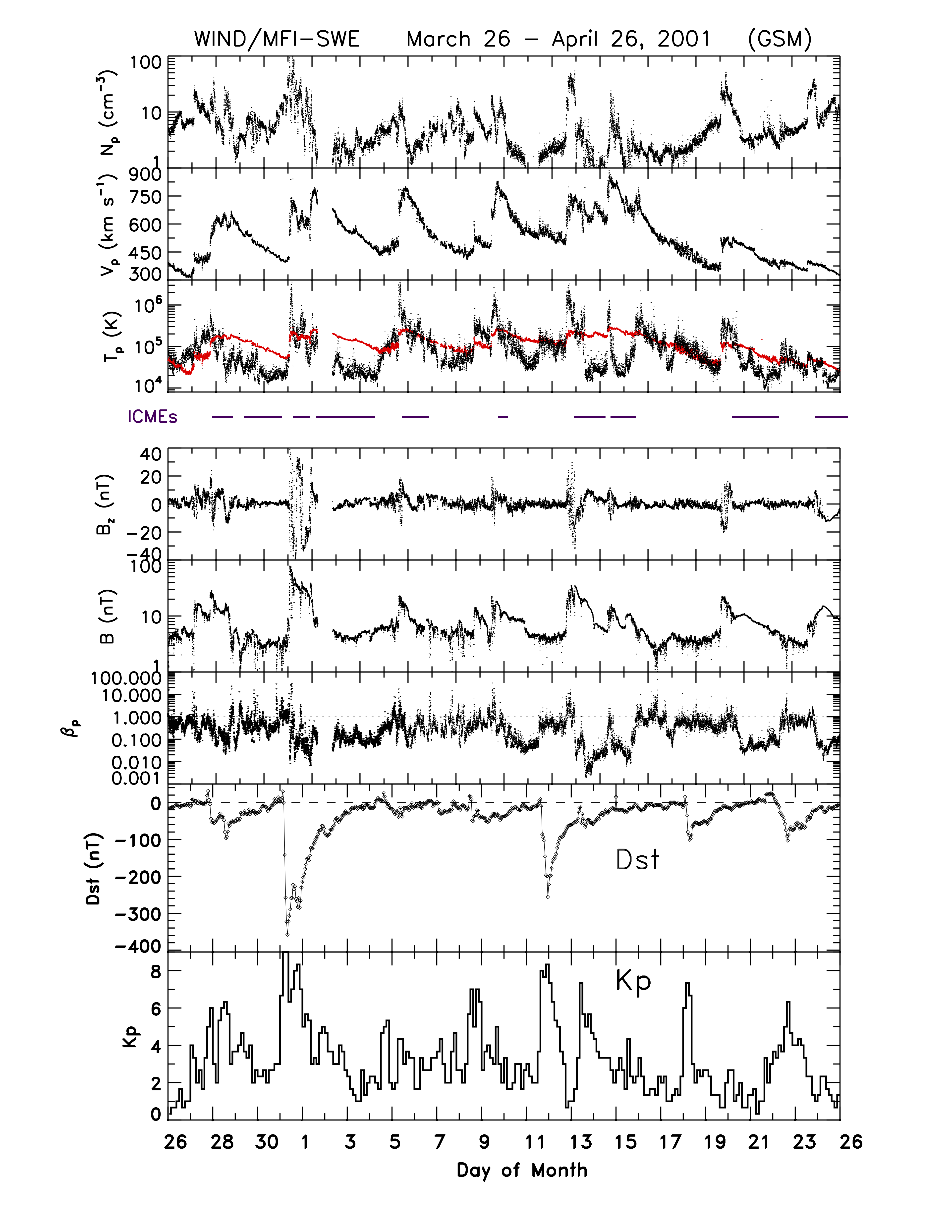}
\label{fig:March}
\vspace{-0.3cm}
\caption{Solar wind measurements and geomagnetic indices of a 31-day period in March-April 2001. The panels show, from top to bottom, the proton density, bulk speed, temperature (expected temperature in red), the eleven ICMEs identified in this
period, the $B_z$ component of the 
magnetic field (in GSM coordinates), the total field, the proton $\beta$, the 
storm-time Dst index and the planetary Kp index.}
\end{figure*}

With so many CMEs, this interval represents a particularly active period at the Sun and in the inner heliosphere (see \opencite{Wang:2003}, \opencite{Berdichevsky:2003}). Correspondingly, the geomagnetic Kp index and the storm-time Dst index indicate that it is also a very disturbed period for the magnetosphere.  The event on 31 March\,--\,1 April 2001 stands out. Here the proton density, the amplitude of the $B_z$ component of the magnetic field, and the magnetic field strength all reach the highest values during this whole interval. The Dst index (uncorrected for magnetopause currents, see below) reaches peak values of $-350$~nT and has a two-pronged profile, and the Kp index saturates (Kp = 9). 

The strong geomagnetic effects of partial or total ejecta mergers may be considered to result from a combination of two effects: on top of the geoeffectiveness of the individual ejecta there is that introduced by the interaction process. Data and simulations have elaborated on aspects of CME-CME interactions which enhance their  geoeffectivness. \inlinecite{Berdichevsky:2003} and \inlinecite{Farrugia:2004} noted the following features: (i) transfer of momentum of the trailing shock and its post-shock flow to the leading CME, (ii) acceleration (deceleration) of the leading (trailing) CME, (iii) strengthening of the shock after its merger with the trailing shock, and (iv) heating and compression of the leading ejecta. 
In simulations, \inlinecite{Lugaz:2005b} emphasized the importance of the trailing shock as it passes through the preceding ejecta to eventually merge with the leading shock, strengthening it. Below, we first give an example of the role of the enhanced density in intensifying the geomagnetic disturbances and then discuss another one where the role of the trailing shock is central.

The geomagnetic storm during the event on 31 March\,--\,1 April 2001 (Figure~10a) has been studied by 
many researchers.  Most relevant here is that \inlinecite{Farrugia:2006} argued that the extreme severity of the storm was ultimately due to a very dense plasma sheet combined with strong southward magnetic fields. They showed that the plasma sheet densities were, in turn, correlated with the high densities in the ejecta merger. Since compression of plasma is a feature of CME-CME interactions, they concluded that the interaction was ultimately responsible for causing this superstorm, with Dst reaching values below $-250$~nT for several hours.

\begin{figure*}[tb]
  \centering
  \includegraphics[width=9cm]{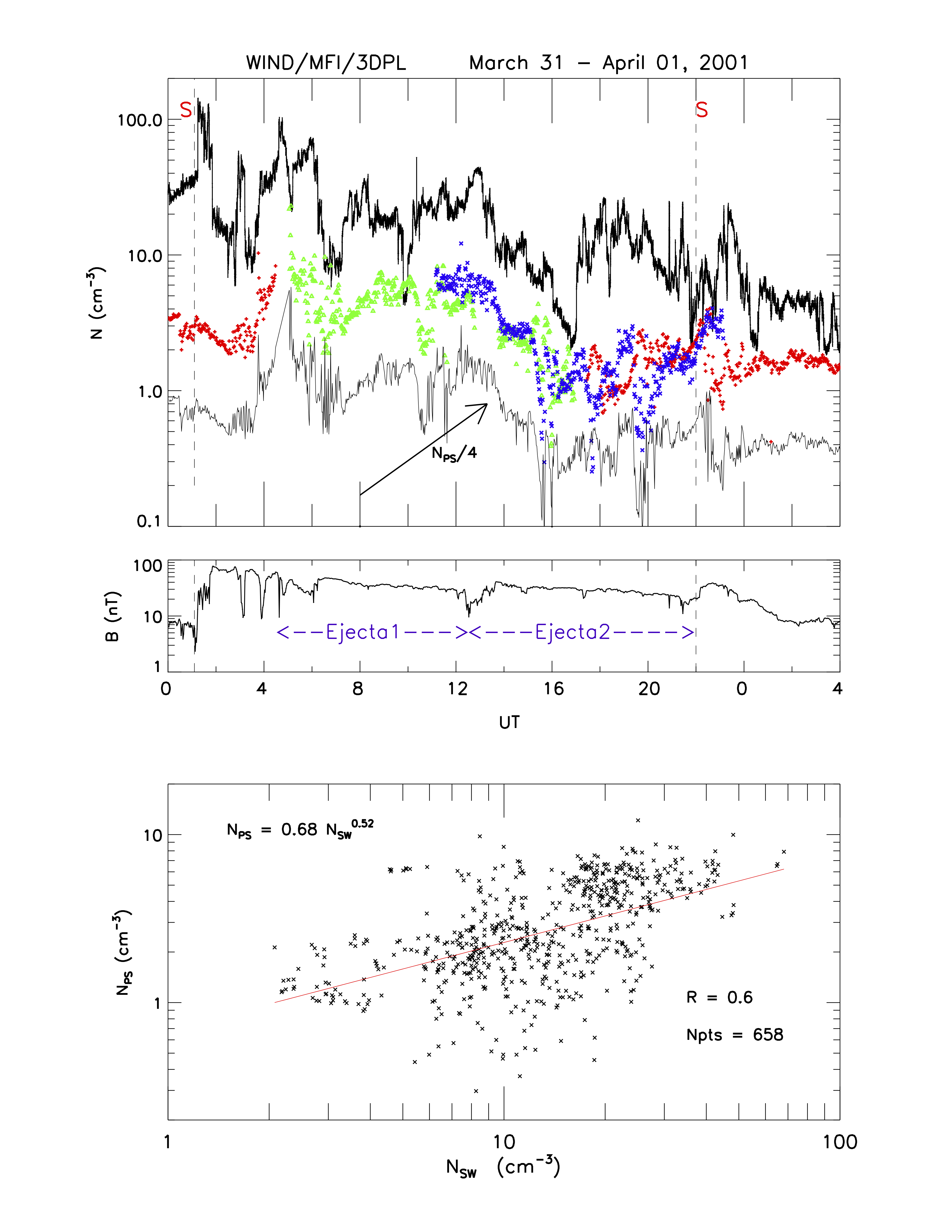}
\label{fig:March_zoom}
\vspace{-0.3cm}
\caption{Comparison of the solar wind and plasma sheet densities during the 31 March\,--\,1 April 2001 CME-CME interaction event. The top panels shows the solar wind density (thick black line), total ion densities above $\sim$100 eV acquired by three {\it Los Alamos National Laboratory} (LANL) spacecraft in geostationary orbit on the nightside in colored symbol, and the compound plasma sheet density (divided by four for clarity). The middle panel shows the total magnetic field. The bottom panel shows a scatter plot of $N_{ps}$ versus $N_{sw}$ and a best fit-line. Adapted from Farrugia {\it et al.} (2006) and published by permission of John Wiley and Sons.}
\end{figure*}

We illustrate this by Figure~12. The middle panel gives the total magnetic field for reference and shows the two interacting ejecta in the process of merging. In the top panel is plotted the density of the interacting ejecta (thick black trace). Below in color is the  density of the plasma sheet obtained from the total ion densities above $\sim$100 eV acquired by three {\it Los Alamos National Laboratory} (LANL) spacecraft in geostationary orbit on the nightside (18-06 MLT). Below, we show this density again as a single-valued function (divided by four for clarity). To produce this, all data were retained where there was no overlap between measurements from the different spacecraft. When there was an overlap,  we kept only data acquired closest to local midnight. One can see that the temporal trends in the densities of plasma sheet ($N_{ps}$) and ejecta ($N_{sw}$) are very similar. The bottom panel shows a scatter plot of $N_{ps}$ versus $N_{sw}$, for which the regression line gives $N_{ps} \sim N_{sw}^{1/2}$.

The strength of the ring current depends mainly on two factors: (i) the electric field in which the particles drift as they travel inwards from the plasma sheet, and (ii) the density of the plasma sheet itself, since it
constitutes the seed population \cite{Jordanova:2006}. In this case, we have a large enhancement of the latter.
Using the global kinetic drift-loss model developed by \inlinecite{Jordanova:1996}, it was shown that the two-dip Dst profile could be reproduced even when the code was run with the  plasma sheet density kept constant at its original value. However, the 
intensity of the storm was thereby grossly underestimated.  It could be adequately 
reproduced only when the plasma sheet density was updated in accordance with the data (see Figure 4 in \opencite{Farrugia:2006}). The end result was a storm where Dst stayed below $-200$~nT for 21 hours and below $-250$~nT  for 7 hours.
In a broader context, \inlinecite{Borovsky:1998} carried out a statistical study relating properties of the plasma sheet with those of the solar wind. One of these was the density, where they found a strong correlation. The relationship they obtained, with $N_{ps}$ scaling as the square root of $N_{sw}$, is similar to that described here. However, compared to their survey, the dynamic range of the density in the case just described was a factor of five higher.

An important implication is that CME-CME interaction can be a viable source of two-dip geomagnetic storms. \inlinecite{Kamide:1998}, studying over 1200 geomagnetic storms, noted that a significant proportion of these were double-dip storms: the Dst index reaches a minimum, recovers for a few hours, and then reaches a second minimum. A traditional view on how this profile comes about is the sheath-ejecta mechanism, that is, the negative $B_z$ fields in the sheath region, which have been compressed by the shock ahead of the ejecta, are responsible for the first Dst drop and the negative $B_z$ phase in the ejecta is then responsible for the second one (\opencite{Tsurutani:1988}, \opencite{Tsurutani:1988}, \opencite{Gonzalez:2002}, and review by
\opencite{Tsurutani:1997}). Here we have given an example of how such two-dip Dst profiles can arise from CME-CME interactions.  In a wide survey of major storms (Dst $< -100$~nT) over Solar Cycle 23 (1996-2006), \inlinecite{Zhang:2008} found that a common source of double-dip storms consists of closely spaced or interacting CMEs. Of course, there are other levels of complexity, such as multiple-dip storms (see example, \opencite{Xiong:2006} and \opencite{Zhang:2008} and \opencite{Richardson:2008}.)

Discussed so far are primarily cases corresponding to multiple-MC events or two (or more) CMEs in close succession but that do not interact. In \inlinecite{Lugaz:2014}, the authors discussed how a long-duration seemingly isolated event resulting from the merging of two CMEs may drive the magnetosphere for a long period, resulting in sawtooth events associated with injection of energetic particles observed at geo-synchronous orbit. In the case of the 19\,--\,22 March 2001 event (see Figure~10b), in addition to sawtooth events, it resulted in an intense geomagnetic storm for which the peak Dst reached $-149$~nT and which remained below the moderate level (below $-50$~nT) for 55 hours.

We now consider a case where the trailing shock passing through the front CME plays a leading role in the resulting geomagnetic disturbances. There are two effects of the interaction on increasing geo-effectiveness. First, the leading ejecta get compressed by either the overtaking shock or magnetic ejecta, or a combination of both \cite{Burlaga:1991,Vandas:1997,Farrugia:2004,Lugaz:2005b,Xiong:2006,Liu:2012}. Assuming conservation of magnetic flux, the compressed southern $B_z$ interval is more geoeffective than uncompressed southern $B_z$ interval (see the quantitative analysis in \opencite{Wang:2003d} and \opencite{Xiong:2007}).
Secondly, the overtaking shock itself and the sheath of compressed ejecta may be more efficient in driving geoeffects. Recently, \inlinecite{Lugaz:2015a} and \inlinecite{Lugaz:2016} published survey studies about  the effect of a shock propagating inside a preceding ejecta on the geo-effectiveness of the shock/sheath region. They found that about half the shocks whose sheaths results in at least a moderate geomagnetic storm are shocks propagating within a previous CME or a series of two shocks. Shocks inside CMEs as a potential source of intense geomagnetic storms were also discussed by \inlinecite{Wang:2003a} and further investigated in \inlinecite{Wang:2003c}. Specific examples were also discussed in \inlinecite{Lugaz:2015b}, who argued that the combination of high dynamic pressure and compressed magnetic field just behind the shock may be particularly efficient in pushing Earth's magnetopause earthwards and, therefore, in driving energetic electron losses in Earth's radiation belt. An example of such an event is given in Figure~10d for a shock inside a magnetic ejecta that occurred on 19 February 2014 and resulted in an intense geomagnetic storm. 

Complex ejecta, due to their complex and turbulent magnetic field and the short duration of any southward periods, often result in strong but not extreme driving of Earth's magnetosphere. For example, the 26\,--\,28 November 2000 event (Figure~10b), caused by a series of homologous CMEs, as discussed in Section~\ref{sec:homologous}, resulted in a peak Dst of $-80$~nT between 26 and 28 November, even though the dawn-to-dusk electric field reached values above 10 mV\,m$^{-1}$, typically associated with intense geomagnetic storms. The large and rapid changes in the orientation of the magnetic field certainly played a role in the reduced geo-effectiveness \cite{Burlaga:2002,Wang:2002,Lugaz:2007}.

The response of the magnetosphere is expected to become nonlinear and even saturate under the strong forcing when interacting ejecta pass Earth, for example, the polar cap potential \cite{Hill:1976,Siscoe:2002} and erosion of the dayside magnetosphere \cite{Muehlbachler:2005}. Furthermore, the correction to the raw Dst index from magnetopause currents (Chapman-Ferraro currents, \opencite{Burton:1975}) becomes inappropriate since the Region 1 currents have taken over the  role of the Chapman-Ferraro currents in standing off the solar wind (\opencite{Siscoe:2005}, and references therein). Following \inlinecite{Vasylinas:2004} and \inlinecite{Siscoe:2005}, one can speak of the magnetosphere as having transitioned from being solar wind-driven to being ionosphere-driven.

\section{Discussions and Conclusion}\label{sec:conclusion}

We have reviewed the main physical phenomena and concepts associated with the initiation and interaction of successive CMEs, including their effects on particle acceleration and their potential for strong driving of Earth's magnetosphere. Initiation mechanisms need to be studied in the photosphere and chromosphere combining observations and simulations. The close relation between CMEs and filaments leads to the urge for more detailed studies on filament evolution and partial eruptions that result in multiple CMEs (twin-CMEs, sympathetic CMEs and homologous CMEs). 

Homologous CMEs and sympathetic CMEs are two main sources of interacting CMEs though unrelated successive CMEs may interact too. Based on the current knowledge, the possible mechanisms of one CME triggering another include (1) destabilization of magnetic structures by removing the overlying field, and (2) the continuous emergence of flux and helicity from the lower atmosphere. However, statistical studies have suggested that homologous or sympathetic CMEs only correspond to a small fraction of total CMEs, meaning that not all of CMEs can trigger subsequent eruptions. It raises the serious question as to why some CMEs trigger another, whereas others do not. Answering this question will directly determine the capability of forecasting of homologous and sympathetic CMEs. Coronal magnetic field is the key information in studying the initiation of CMEs. However, it is now mostly known from extrapolation methods based on limited observations. {\it Solar Orbiter}, to be launched in 2018, will provide vector magnetic field at the photosphere from a point of view other than that of the Earth. Combined with the magnetic field observed by SDO, it may increase the accuracy of the coronal magnetic field extrapolation. 

With respect to particle acceleration, it is clear that SEPs are strongly affected by preconditioning in terms of turbulence and seed population. Radio enhancements, themselves, are likely to be directly related to the CME-CME interaction process. SEPs, on the other hand, are more likely to be related to having a succession of (not necessarily interacting) CMEs. Enhanced levels of turbulence and suprathermals following a first CME are likely physical explanations. Studies have shown that type II radio burst intensification is due to the front of a faster, second CME colliding with the rear of a slower, preceding CME; however, shock-streamer interaction is also a likely cause of such enhancements. In fact, enhanced type II radio signatures may occur for different types of interaction scenarios.

Wide-angle heliospheric imaging of CMEs, now routinely performed by STEREO, has resulted in a new era of study of CME-CME interaction where 3D kinematics can be combined with Sun-to-Earth imaging of succession of CMEs, 3D numerical simulations and radio observations. Many of the recent studies have focused on the relation between the interaction and the resulting structure measured at 1 AU as well as the exchange of momentum between the CMEs. In many instances, after their interaction, the two CMEs are found to propagate with a uniform speed profile, similar to that of an isolated CME, but how this final speed relates to that of the two CMEs before the interaction is still an area of active research. The ``compression'' of the leading CME (associated with a reduced rate of expansion) and the potential deflection of the CMEs are two additional effects of CME-CME interaction that can have a strong influence on space weather forecasting. The exchange of momentum between CMEs is likely to involve a series of phases, including compression of the leading CME by the trailing shock wave that increases the CME magnetic energy, followed by an over-expansion of the leading CME, and potential reconnection between the two magnetic ejecta.

Most studies still primarily focus on the different aspects of CME-CME interaction (initiation, SEP acceleration, heliospheric interaction, {\it in situ} measurements, geo-effects) without attempting to obtain a global view. One issue is that heliospheric imaging provide information about different density structures and their kinematics but not about the magnetic field. 
More global studies, for example, to determine whether sympathetic or homologous (or unrelated) eruptions are more likely to result in CME-CME interaction, and to which structures at 1~AU they correspond are now possible with modern remote imaging but have not been undertaken yet. Study of SEP acceleration and radio bursts associated with successive CMEs must be combined with study of the stereoscopic observations to determine the de-projected CME directions, distances and kinematics, as done only for few events so far \cite{Temmer:2014,Ding:2014}. Similar studies may help to investigate which time delay between successive CMEs result in the most intense geo-effects, and what the most extreme scenario can be. A recent investigation by the Cambridge Centre for Risk Studies of the economical consequences of an extreme CME took as its base scenario the impact in close succession of two non-interacting CMEs, similar to the perfect storm scenario of \inlinecite{Liu:2014}. It is as yet unclear if such a scenario would result in stronger magnetospheric, ionospheric and ground currents as compared to two CMEs in the process of interacting (shock inside CME) or having interacted (multiple-MC event). 

Although detailed investigations of CME-CME interaction became more frequent following the report by \inlinecite{Gopalswamy:2001} of a total disappearance of one CME following its overtaking by a faster CME (referred to as CME ``cannibalism''), total magnetic reconnection might not take place regularly, as it might be too slow for effectively merging entire flux ropes. For this reason, the most common result of CME-CME interaction at 1~AU is multiple-MC events, where the individual ejection can be distinguished, complex ejecta where some of the individual characteristics are lost and shocks propagating inside a previous CME. Each of these has a different way to interact with Earth's magnetosphere. Often, the long-duration driving and compressed magnetic fields result in intense geo-effects; in addition shocks within CME have a higher probability of having a geo-effective sheath as compared to shocks propagating into typical solar wind conditions.

Currently available {\it in situ} data show us only the consequences of CME-CME interaction but not the interaction process itself.  {\it In situ} data during the interaction process is needed, as was possible with Helios \cite{Farrugia:2004}. {\it Solar Orbiter} and {\it Solar Probe+} will provide opportunities for measuring the interaction process as it occurs, and to determine the fate of shocks propagating inside ejecta, the irreversibility (or not) of the interaction. Studies made possible by these spacecraft when they are in alignment (conjunction) will help us to answer some of these outstanding questions.

 \begin{acks}
We acknowledge the following grants: NASA grant NNX15AB87G and NSF grants AGS1435785, AGS1433213 and AGS1460179 (N.~L.), NSFC grants 41131065 and 41574165 (Y. W.), Austrian Science Fund FWF: V195-N16 (M.~T.), and NASA grant NNX16AO04G (C.~J.~F.). N.~L. would like to acknowledge W.~B. Manchester, I.~I. Roussev, Y.~D. Liu, G. Li, J.~A. Davies, T.~A. Howard, and N.~A. Schwadron for fruitful discussion about CME-CME interaction over the years, as well as the VarSITI program. Some of the Figures in this article have been published by permission of the AAS and John Wiley and Sons as indicated in the text.

Disclosure of Potential Conflicts of Interest: The authors declare that they have no conflicts of interest.
\end{acks}

\bibliographystyle{spr-mp-sola-cnd}

\end{article}

\end{document}